\documentclass[11pt,final]{siamltex} 

\usepackage{mathrsfs}
 
\usepackage{graphicx}
\usepackage{amsfonts}
\usepackage{amsmath}
\usepackage{amssymb}
\usepackage{mathtools}
\usepackage{bm}
\usepackage{amsbsy}
\usepackage{algorithmic}
\usepackage{algorithm}
\usepackage{subfigure}
\usepackage{listings}
\usepackage{tikz}
\usepackage{stackrel}
\usepackage{url}
\usepackage{xspace}
\usepackage{enumerate}
\usepackage{enumitem}
\usepackage{soul,xcolor} %
\setstcolor{red} %

\usepackage{color}            %

\usepackage{booktabs}
\usepackage{siunitx}
\newcommand*\mc[1]{\multicolumn{1}{c}{#1}}

\usepackage[retainorgcmds]{IEEEtrantools}

\usepackage[T1]{fontenc}
\usepackage[ansinew]{inputenc}
\usepackage{lmodern}

\usepackage{ifpdf}
\ifpdf
  \usepackage{epstopdf} 
\fi

\usepackage[letterpaper,centering]{geometry}
\usepackage[colorlinks=true, citecolor=blue, filecolor=black,linkcolor=red, urlcolor=blue, hypertexnames=false]{hyperref}

\usepackage[textsize=tiny]{todonotes}

\newcommand{\hblue}{\textcolor{black} }
\newcommand{\hrevone}{\textcolor{black} }

\newcommand{\eb}{\boldsymbol{e}}

\newcommand{\fb}{\boldsymbol{f}}

\newcommand{\sbo}{\boldsymbol{s}}

\newcommand{\xb}{\boldsymbol{x}}

\newcommand{\Xb}{\boldsymbol{X}}
\newcommand{\Yb}{\boldsymbol{Y}}
\newcommand{\Zb}{\boldsymbol{Z}}

\newcommand{\Bc}{{\cal B}}

\newcommand{\Mc}{{\cal M}}
\newcommand{\Nc}{{\cal N}}

\newcommand{\Sc}{{\cal S}}

\newcommand{\Ex}{\mathbb{E}}

\newcommand{\Var}{{\mathbb{V}{\rm ar}}}

\def\zal2{z_{\alpha/2}}

\newcommand{\Gy}{\Gamma_{ \Yb}}

\newcommand{\Gobs}{\Gamma_{\rm obs}}
\newcommand{\Gpr}{\Gamma_{\rm pr}}
\newcommand{\Gauss}{\Nc}

\newcommand{\Gprs}{S_{\rm pr}} %

\newcommand{\Gposhs}{\widehat{S}_{\rm pos}}

\newcommand{\Gobss}{S_{\rm obs}}

\newcommand{\Gpos}{\Gamma_{\rm pos}}
\newcommand{\Gz}{\Gamma_{\Zb}}

\newcommand{\Gzyh}{\widehat{\Gamma}_{\Zb|\Yb}}

\newcommand{\Gzysub}{\widehat{\Gamma}_{\Zb|\Yb}}

\newcommand{\Gzyopt}{\widetilde{\Gamma}_{\Zb|\Yb}}
\newcommand{\Gzyopts}{\widetilde{S}_{\Zb|\Yb}}

\newcommand{\Gzy}{ \Gamma _{\Zb|\Yb}}

\newcommand{\nuz}{\nu_{\Zb}}

\newcommand{\Gposh}{\widehat{\Gamma}_{\rm pos}}
\newcommand{\mupos}{\mu_{\rm pos}(\Yb)}
\newcommand{\muzy}{\mu_{\Zb|\Yb}(\Yb)}
\newcommand{\muzyapp}{\widetilde{\mu}_{\Zb|\Yb}(\Yb)}

\newcommand{\wq}{\widehat{q}}

\newcommand{\mzy}{\nu_{\Zb \vert \Yb } }
 \newcommand{\mzya}{\tilde{\nu}_{\Zb \vert \Yb } }
\newcommand{\pizy}{\pi_{\Zb \vert \Yb } }

 \newcommand{\pizya}{\tilde{\pi}_{\Zb \vert \Yb } }

\newcommand{\re}{\mathbb{R} }
\newcommand{\ra}{\rightarrow }

\newcommand{\Eb}{ \mathbb{E} }

\newcommand{\gop}{\mathcal{O}} 
\newcommand{\gopd}{\gop_\dagger} 
\newcommand{\Dom}{\mathcal{D}} 
\newcommand{\Temp}{\Theta} 
\newcommand{\Tempb}{\mathbf{\Temp}} 
\newcommand{\Tempbh}{\mathbf{\Temp}_h} 
\newcommand{\Tempbhz}{\mathbf{\Temp}_{0h}} 

\newcommand{\nop}{\mathscr{J}}

\newcommand{\mupr}{\mu_\text{pr}} 
\newcommand{\error}{\boldsymbol{\mathcal{E}}} 
\newcommand{\errorTr}{\boldsymbol{\Delta} }

\newcommand{\Zbh}{\widehat{\Zb}} 
\newcommand{\dr}{d_{\mathcal{R}}} 
\newcommand{\spdman}{\it{Sym}^+ }

\title{\vspace{-10pt}Goal-oriented optimal approximations of Bayesian
  linear inverse problems 
}

\author{Alessio Spantini\footnotemark[1] \and Tiangang Cui\footnotemark[2]
\and Karen Willcox\footnotemark[1]\,\,,\\ 
Luis Tenorio\footnotemark[3] \and Youssef
Marzouk\footnotemark[1]}

\begin{document}

\maketitle

\renewcommand{\thefootnote}{\fnsymbol{footnote}}

\footnotetext[1]{Department of Aeronautics and Astronautics, Massachusetts Institute of Technology, Cambridge, MA
02139, USA, \texttt{\{spantini, kwillcox, ymarz\}@mit.edu}.}

\footnotetext[2]{School of Mathematical Sciences, Monash University, Victoria 3800, Australia, 
\texttt{tiangang.cui@monash.edu}.}

\footnotetext[3]{Applied Mathematics and Statistics, Colorado School of Mines, Golden, CO 80401, USA, \texttt{ltenorio@mines.edu}.}

\renewcommand{\thefootnote}{\arabic{footnote}}

\begin{abstract} 
  We propose optimal dimensionality reduction techniques for the solution of
  goal--oriented linear--Gaussian inverse problems, where the quantity of
  interest (QoI) is a function of the inversion parameters.  These
  approximations are suitable for large-scale applications.  In particular, we
  study the approximation of the posterior covariance of the QoI as a low-rank
  negative update of its prior covariance, and prove optimality of this update
  with respect to the natural geodesic distance on the manifold of symmetric 
  positive definite matrices.  Assuming exact knowledge of the posterior mean of
  the QoI, the optimality results extend to optimality in distribution with
  respect to the Kullback-Leibler divergence and the Hellinger distance between
  the associated distributions. We also propose the approximation of the posterior
  mean of the QoI as a low-rank linear function of the data, and prove
  optimality of this approximation with respect to a weighted Bayes risk.  Both
  of these optimal approximations avoid the explicit computation of the full
  posterior distribution of the parameters and instead focus on directions that
  are well informed by the data \textit{and} relevant to the QoI. 
  These directions stem from a balance among all the components of the
  goal--oriented inverse problem: prior information, forward model,
  measurement noise, and ultimate goals.  We illustrate the theory using a
  high-dimensional inverse problem in heat transfer.
\end{abstract}



\pagestyle{myheadings}
\thispagestyle{plain}

\markboth{\MakeUppercase{Spantini et al.}}
{\MakeUppercase{Goal--oriented inverse problems}}

\section{Introduction}
\label{s:intro}
The Bayesian approach to inverse problems treats the unknown parameters as random
variables, endowed with a prior distribution that encodes one's knowledge before data are
collected. The distribution of the data conditioned on any value of the parameters is
specified through the likelihood model.
Bayes' rule then combines prior and likelihood information to yield the posterior
distribution, i.e., the distribution of the parameters conditioned on the data.  The
posterior distribution defines the Bayesian solution to the inverse problem.
Characterizing this posterior distribution is of primary interest in many engineering and
science applications, ranging from computerized tomography and optical imaging
to geostatistical modeling.
For instance, we might be interested in the posterior marginals, the posterior probability
of some functionals of the parameters, or the probability of rare events under the
posterior measure.  In all these cases we may need to draw samples
from the posterior distribution.  This sampling task tends to be
extremely challenging in large-scale applications, especially when the
parameters represent a finite-dimensional approximation to a
distributed stochastic process like a permeability or a temperature
field.  In many applications, however, we are only interested in a
particular \textit{function} of the parameters (e.g., the temperature
field over a subregion of the entire domain or the probability that
the temperature exceeds a critical value).  In this paper we exploit
such \textit{ultimate goals} to reduce the cost of
inversion.  %

We begin by considering  a finite-dimensional linear-Gaussian inverse problem of the form 
	\begin{equation} \label{eq:linearModelParam}
		\Yb = G\,\Xb + \error,
	\end{equation}
where $\Xb\in\re^n$ represents the unknown parameters, $\Yb \in \re^d$ denotes the noisy observations, $G\in\re^{d \times n}$ is a 
\hrevone{deterministic} linear forward operator,  and  $\error \sim \Gauss(0,\Gobs)$ is a
zero-mean additive Gaussian noise, statistically independent of $\Xb$ and with covariance matrix $\Gobs \succ 0$.
\hrevone{
(We use boldface capital letters to denote random vectors.)}
We prescribe a Gaussian prior distribution, $\Gauss(0,\Gpr)$, on 
$\Xb$ and assume, without loss of generality,  zero prior mean and  
$\Gpr\succ 0$.
One is usually concerned with the posterior distribution of the parameters,
$\Xb \vert \Yb \sim \Gauss( \mupos, \Gpos )$,\footnote{$\Xb \vert \Yb$ refers to a random variable distributed according to the measure of $\Xb$ conditioned on   
$\Yb$.
} which has mean and covariance given by 
\begin{equation} \label{eq:statistics}
	\mupos = \Gpos\, G^\top \Gobs^{-1}\, \Yb, \qquad \Gpos = ( H + \Gpr^{-1} )^{-1},
\end{equation} 
where $H\coloneqq G^\top \Gobs^{-1} \,G$ is the Hessian of the negative log-likelihood.
In this paper, however,  we are not interested in the parameters $\Xb$ per se, but rather in a quantity of interest (QoI) $\Zb$ that is a function of the parameters,
\begin{equation} \label{eq:linearModelQoI}
  \Zb = \gop \Xb,
\end{equation}
for some linear and, without loss of generality, full row-rank operator
$\gop\in\re^{p \times n}$ with $p<n$. Our interests are thus \textit{goal-oriented}, as we
wish to characterize only $\Zb$ and not the parameters $\Xb$. Including such ultimate
goals in the inference formulation is an important modeling step in many applications of
Bayesian inverse problems. This additional step should reduce the computational complexity
of inference by making the ultimate goals explicit. Nevertheless, it is still not clear
how to leverage ultimate goals to yield more efficient Bayesian inference algorithms. The
present paper will address this issue.

\medskip

The full Bayesian solution to the goal-oriented inverse problem is the posterior distribution of
the QoI, i.e., $\Zb|\Yb$.	
It is easy to see that $\Zb|\Yb$ is once again Gaussian with mean and covariance matrix given by
\begin{equation} \label{eq:QoIstatistics}
	\muzy = \gop \,\mupos, \qquad \Gzy = \gop \, \Gpos \, \gop^\top.
\end{equation}
The goal of this paper is to characterize statistically optimal, 
 computationally efficient, and structure--exploiting approximations of the  statistics of $\Zb \vert \Yb$ whenever
the use of direct formulas such as \eqref{eq:QoIstatistics} is challenging or impractical 
(perhaps due to high computational complexity or excessive storage requirements). 
 We will approximate $\Gzy$ as a low-rank negative update of the prior covariance of the QoI. Optimality will be defined with respect to the natural geodesic distance on the manifold of symmetric and positive definite (SPD) matrices \cite{forstner2003metric}. 
The posterior mean $\muzy$ will be approximated as a low-rank function of the data, where optimality
is defined by the minimization of the Bayes risk for  
squared-error  loss weighted by $\Gzy^{-1}$.
The essence of these approximations is the restriction of  the inference process to directions in the parameter space that are  
informed by the data relative to the prior \textit{and} that are relevant to the QoI.
These directions correspond to the leading generalized eigenpairs of a
suitable matrix pencil.

This paper extends, in several different ways, the work on goal-oriented inference
originally presented in \cite{lieberman2012}. First of all, we introduce the notion of
\textit{optimal approximation}, rather than exact computation, for both the posterior
covariance matrix and the posterior mean of the QoI.  We propose computationally
efficient algorithms to determine these optimal approximations.  The complexity of our
algorithms scales with the intrinsic dimensionality of the goal-oriented problem---which
here reflects the dimension of the parameter subspace that is simultaneously relevant to
the QoI and informed by the data,
as noted above.
In particular, the full posterior distribution of the parameters need not be computed at
any stage of the algorithms. This is a key
contrast with \cite{lieberman2012}.  Moreover,
we make it possible to handle high-dimensional QoIs such as those arising from the
discretization of a distributed stochastic process.  This class of problems is frequently
encountered in applications (see, e.g., Section \ref{s:examples}). 

The ideas and algorithms of this paper are primarily developed in the
linear-Gaussian case. On one hand, their application to goal-oriented linear inverse
problems is of standalone interest \cite{lieberman2012}.  On the other hand, in the
context of high-dimensional nonlinear Bayesian inverse problems, the Gaussian
approximation is often the only tractable approximation of the posterior distribution
\cite{bui2012extreme, isaac2015scalable}.  For example, in \cite{isaac2015scalable}, a linearization of
the parameter-to-observable map and a linearization of the parameter-to-prediction map are
performed for a high-dimensional ice sheet model, reducing the nonlinear inverse problem
to a goal-oriented linear Gaussian inverse problem.  Moreover, the rigorous analysis of
dimensionality reduction ideas in linear inverse problems often leads to computationally
efficient dimensionality reduction strategies for 
nonlinear inverse \hrevone{problems}\footnote{\hrevone{Of course, a comprehensive theory for nonlinear inverse problems 
will likely require tools beyond those explored in this paper.}
}
(see, e.g.,
\cite{spantini2014optimal,cui2014likelihood} or
\cite{lieberman2012,lieberman2014nonlinear}).

The remainder of the paper is organized as follows.  In Section \ref{s:theory} we
introduce statistically optimal approximations of the posterior statistics of the QoI.
\hrevone{ 
Section \ref{s:simpleExample} contains a simple proof-of-concept example, while
in Section \ref{s:examples} we illustrate the theory using a more realistic
inverse problem in heat
transfer.
Section \ref{s:conclusions} offers some concluding remarks.
Appendix \ref{sec:diffGeom} reviews an important class of metrics
between distributions (Rao's distance), while
Appendix
\ref{sec:proofs} contains the proofs of the main results of this paper.
}  
\section{Theory}
\label{s:theory}
In this section we introduce optimal approximations of the posterior
mean $\muzy$ and posterior covariance $\Gzy$ of the QoI.  Section
\ref{s:metrics} reviews a class of metrics between probability
distributions and introduces natural loss functions for the
approximation of $\muzy$ and $\Gzy$.  Section \ref{sec:approx_cov_QoI}
then focuses on the approximation of $\Gzy$, while the posterior mean
approximation is addressed in Section \ref{sec:approx_mean_QoI}.  The
main results of this section are Theorem \ref{approxCovUpdate-qoi} and
Theorem \ref{thm:mean_approx_lowrank}.

\subsection{Optimality criteria: metrics between distributions}
\label{s:metrics}
To measure the quality of the approximation of a posterior distribution 
we employ a metric first introduced by Rao in \cite{rao1945info} based on the Fisher information.
Rao's approach to comparing distributions is rooted in differential geometry.
The idea is to turn a parametric family of distributions into a Riemannian manifold endowed
with  a metric based on the geodesic distance \cite{rao1949distance}.
The  Riemannian structure  is induced by a quadratic form defined by the Fisher information 
matrix. 
\hrevone{
(See Appendix \ref{sec:diffGeom} for the explicit construction of Rao's distance.)}
For a Gaussian family of distributions  this metric
can be written explicitly 
at least in two 
particular cases \cite{atkinson1981rao,skovgaard1984riemannian}. 
If the family consists of Gaussian distributions with the same covariance matrix, $\Gamma$, then
the metric reduces to the Mahalanobis distance between the means \cite{mahalanobis1936generalized,rao1949distance}:
\begin{equation}
	\dr( \nu_1 , \nu_2 ) = \| \mu_1- \mu_2 \|_{\Gamma^{-1}}, \qquad \,\nu_1=\Gauss(\mu_1 , \Gamma),\,
	\nu_2=\Gauss(\mu_2 , \Gamma),
\end{equation}
where $ \| z \|_{\Gamma^{-1}}^2 \coloneqq z^\top \Gamma^{-1} z$.
We will use this metric to define optimal approximations of the posterior mean in Section
\ref{sec:approx_mean_QoI}. 
Notice that this metric emphasizes differences in the mean along eigendirections
of $\Gamma$ corresponding to low variance.

If, on the other hand, the family consists of Gaussian distributions with the same mean, $\mu$,
then the metric reduces to:
\begin{equation} \label{eq:distCov}
	\dr( \nu_1 , \nu_2 ) = \sqrt{ 
	\frac{1}{2}\sum_i \ln^2(\sigma_i)}, \qquad \,\nu_1=\Gauss(\mu , \Gamma_1),\,
	\nu_2=\Gauss(\mu, \Gamma_2),
\end{equation}
where $(\sigma_i)$ are the generalized eigenvalues of the pencil $(\Gamma_1,\Gamma_2)$
\cite{jensen1976}.
That is, $(\sigma_i)$ are the roots of the characteristic
polynomial $\det(\Gamma_1-\sigma\,\Gamma_2)$ and
satisfy the equation $\Gamma_1\,v_i = \Gamma_2\,v_i\,\sigma_i$ for
a collection of generalized eigenvectors $(v_i)$ \cite{golub2012matrix}.
Since this family of distributions can be identified with the cone of 
SPD matrices,
$\spdman$,
\eqref{eq:distCov} can also be used as a Riemannian metric on $\spdman$ 
\cite{forstner2003metric,bhatia2009positive}.
We will use this Riemannian metric  to define optimal approximations of the posterior covariance matrix 
in Section
\ref{sec:approx_cov_QoI}. 
This metric on $\spdman$ is also the unique geodesic distance that 
satisfies the following two important invariance properties:
	\begin{equation} \label{eq:invarProp}
		\dr(A,B) = \dr(A^{-1},B^{-1}) \quad \text{and} \quad  \dr(A,B) = 
		\dr(MAM^\top,MBM^\top),
	\end{equation}
for any nonsingular matrix $M$ and matrices $A,B\in \spdman$ (e.g., \cite{bonnabel2009riemannian})
making it an ideal candidate to compare covariance matrices.
Moreover, it has
been used successfully in a variety of 
applications (e.g., \cite{smith2005covariance,pennec2006riemannian,moakher2011riemannian,barachant2013classification,horev2015geometry,fletcher2004principal}).
Notice that the {\it flat} distance induced by the Frobenius norm does not satisfy the 
invariance properties  \eqref{eq:invarProp}, and has often been shown to be inadequate for comparing
covariance matrices \cite{fletcher2004principal,arsigny2007geometric,sommer2010manifold}.

In the most general case of manifolds of Gaussian families
parameterized by both the mean and covariance, there seems to be no
explicit form for the geodesic distance.

\subsection{Approximation of the posterior covariance of the QoI}
\label{sec:approx_cov_QoI}
We first focus on approximating $\Gzy$. 
The
cost of computing $\Gzy$ according to \eqref{eq:QoIstatistics} is dominated by the
solution of $p$ linear systems with coefficient matrix
 $\Gpos^{-1}$ in order to
determine $\Gpos \, \gop^\top$.  
\hrevone{
In large-scale inverse
  problems only the action of the precision matrix $\Gpos^{-1}$ on a vector is usually
  available; it is not reasonable to expect to have direct factorizations of
  $\Gpos^{-1}$, such as Cholesky decomposition.  Thus, the solution of
  linear systems with coefficient matrix $\Gpos^{-1}$ is often necessarily iterative
  (e.g., via %
\hrevone{Krylov subspace methods  for SPD matrices} \cite{hestenes1952methods,golub2012matrix}).
} Moreover, 
the storage requirements for $\Gzy$ scale as
$O(p^2)$.  If the dimension of the QoI is inherently low, e.g., $p=O(1)$, then the use of
direct formulas like \eqref{eq:QoIstatistics} can be remarkably efficient.  For instance,
if we are only interested in the average of $\Xb$, i.e.,
$\Zb\coloneqq \frac{1}{n}\sum_i X_i$, then the QoI is only one-dimensional and computing
the posterior covariance of the QoI amounts to solving essentially a single linear system.
As the dimension of the QoI increases, however, direct formulas like
\eqref{eq:QoIstatistics} become increasingly impractical due to high computational
complexity and storage requirements. In many cases of interest, the dimension of
the QoI can be arbitrarily large.  Consider the following simple example. If $\Xb$
represents a finite-dimensional approximation of a spatially distributed stochastic
process (e.g., an unknown temperature field), then the QoI could be the restriction of
this process to a domain of interest.  In this case, the QoI is also a finite-dimensional
approximation of a spatially distributed process, and its dimension can be
arbitrarily high depending on the chosen level of discretization. (We will revisit this example
in Section \ref{s:examples}.)  There is a clear need for new inference algorithms to
efficiently tackle such problems.

\subsubsection{Background on optimal low-rank approximations}
\label{sec:background}
Even though direct formulas like \eqref{eq:QoIstatistics} can be intractable when the QoI
is high-dimensional, essential features of large-scale Bayesian inverse problems bring
additional structure to the Bayesian update: The prior distribution 
\hrevone{might} encode some
kind of smoothness or correlation among the inversion parameters.  Observations are
typically limited in number, indirect, corrupted by noise, and related to the inversion
parameters by the action of a forward operator that filters out some information
\cite{spantini2014optimal,cui2014likelihood}.  As a result, data are usually informative,
relative to the prior, only about a low-dimensional subspace of the parameter space.  That
is, the important differences between the prior and posterior distributions are confined
to a low-dimensional subspace.  This source of low-dimensional structure is key to the
development of efficient Bayesian inversion algorithms
\cite{flath2011fast,cui2014likelihood} and also plays a crucial role when dealing with
goal-oriented problems.

In \cite{spantini2014optimal} we studied the optimal approximation of the posterior
covariance of the parameters as a negative definite low-rank update of the prior
covariance matrix. Optimality was defined with respect to the Riemannian metric given in
Section \ref{s:metrics}, i.e., the geodesic distance on the manifold of SPD matrices
\cite{forstner2003metric}.  Note that optimality of the covariance approximation in this
metric also leads to optimality of the approximation with respect to familiar
measures of similarity between probability distributions, such as the Kullback--Leibler
divergence and the Hellinger distance \cite{spantini2014optimal,pardo2005statistical}.  In
particular, we focused on the approximation class
	\begin{equation} \label{eq:classLinear}
		\Mc_r = \{ \Gpr - KK^\top \succ 0 : \rank(K)\le r \}
	\end{equation}
of positive definite matrices that can be written as a low-rank update of the prior
covariance. This class takes advantage of the low-dimensional structure of the prior-to-posterior
update.\footnote{%
Many approximate inference algorithms, especially in the context of Kalman filtering, exploit the class \eqref{eq:classLinear} to deliver an approximation of $\Gpos$ 
(e.g., \cite{auvinen2009large,auvinen2010variational,solonen2012variational}). These
algorithms, however, are suboptimal in the sense defined by the forthcoming Theorem
\ref{thm:mainLinear}. See \cite{spantini2014optimal} for further details and numerical examples.
}	
The following theorem from \cite{spantini2014optimal} characterizes the optimal
approximation of $\Gpos$, and is the launching point for this section.  Henceforth
whenever we write that $(\alpha, v)$ are eigenpairs of $(A,B)$ we mean that
$(\alpha, v)$ are the generalized eigenvalue--eigenvector pairs of the matrix pencil
$(A,B)$.
\medskip

\begin{theorem}[Optimal posterior covariance approximation]
\label{thm:mainLinear}
Let $(\delta_i^2,w_i)$ be the eigenpairs of 
$(H,\Gpr^{-1})$ with the ordering $\delta_i^2\ge\delta_{i+1}^2$, where
$H \coloneqq G^\top \Gobs^{-1} \,G$  as in \eqref{eq:statistics}. 
Then a minimizer $\Gposh$ of the Riemannian metric $\dr$ between $\Gpos$ and an element of $\Mc_r$ is given by
			\begin{equation} \label{eq:optCovLinear}
				\Gposh = \Gpr - KK^\top, \qquad KK^\top = \sum_{i=1}^r 
				\frac{\delta_i^2}{1+\delta_i^2} \, w_i \,w_i^\top,
			\end{equation}
 where the distance between $\Gpos$ and the optimal approximation is
			\begin{equation} \label{eq:distRlinearpaper}
				\dr^2(\Gpos,\Gposh) =   
				\frac{1}{2}	\sum_{i>r} \,\ln^2(1+\delta_i^2).
			\end{equation}
\end{theorem}

Theorem \ref{thm:mainLinear} shows that the optimal way to update the prior covariance matrix to obtain an approximation of $\Gpos$ is along the eigenvectors of 
 $(H,\Gpr^{-1})$.\footnote{%
The properties of the pencil $(H,\Gpr^{-1})$ have been studied extensively in the literature on classical regularization techniques for linear inverse problems
(e.g., \cite{hansen1989regularization,dykes2014simplified,calvetti2005priorconditioners,	  van1976generalizing,paige1981towards}).
These papers, however, do not adopt a statistical approach to inversion and thus have not considered the optimal approximation of the posterior covariance matrix. 
}
 These eigenvectors are the directions most informed by the data, and are obtained from a  
balance between 
the forward model, the measurement noise, and prior information.
This update is typically low-rank for precisely the reasons discussed above:
the data are informative relative to the prior only about a low-dimensional subspace of the parameter space \cite{bui2012analysis,schillings2014scaling}.
Notice that \eqref{eq:optCovLinear} is not only an optimal and
structure-exploiting approximation of $\Gpos$, but it is also computationally efficient, as the dominant eigenpairs of
$(H,\Gpr^{-1})$ can easily be computed using matrix-free algorithms like a Lanczos iteration (including its block version)
\cite{lanczos1950iteration,paige1972computational,
cullum1974block,bui2012extreme,kalmikov2014hessian} or a randomized SVD
\cite{halko2011finding, liberty2007randomized}.
This approximation of $\Gpos$, originally introduced in \cite{flath2011fast} for
computational convenience and justified by intuitive arguments, has been employed
successfully in many large-scale applications of Bayesian inversion 
\cite{bui2012extreme,martin2012stochastic,
bui2012adaptive,cui2014likelihood,clm_2014,petra2014computational}.
It is the starting point for our analysis of goal-oriented inverse problems.

\subsubsection{A na\"{i}ve approximation}
\label{sec:naiveApprox}
\hblue{In this section we introduce an
intuitive but suboptimal approximation of $\Gzy$, which will provide insight and
motivation for the structure of the forthcoming optimal approximation.
The reader interested exclusively in the optimal approximation may jump directly 
to Section \ref{sec:optApproxCovQoI}.}

The combination of Theorem \ref{thm:mainLinear} with the direct formulas 
\eqref{eq:QoIstatistics}  suggests a first approximation strategy for the posterior
covariance of the QoI:
just replace $\Gpos$ in \eqref{eq:QoIstatistics}  with the optimal approximation
described by Theorem \ref{thm:mainLinear},
\begin{equation} \label{eq:approxCovQoIsub}
	\Gzy \approx \Gzysub \coloneqq \gop \, \Gposh \, \gop^\top = \gop \, \Gpr \, \gop^\top - 
	\gop \, KK^\top \, \gop^\top,
\end{equation}
where the low-rank update $KK^\top$ is given by \eqref{eq:optCovLinear}.
Approximation \eqref{eq:approxCovQoIsub} is already a major computational 
improvement over the direct formulas 
\eqref{eq:QoIstatistics}. There is no need to solve $p$ linear systems; rather, we only need to compute the leading eigenpairs of $(H,\Gpr^{-1})$ with a matrix-free algorithm. 
The rank of the update depends on the dimension of the parameter subspace
that is most informed by the data. 

Despite these favorable computational properties, the approximation
\eqref{eq:approxCovQoIsub} is still not satisfactory as it does not explicitly account for
the goal-oriented feature of the problem: $\Gposh$ in \eqref{eq:approxCovQoIsub} is the
optimal approximation of the posterior covariance of the parameters, but is by no means
tailored to the QoI.  The pencil $(H,\Gpr^{-1})$ used to compute the approximation
$\Gposh$ does not include the goal-oriented operator $\gop$.  In other words, the directions
$(w_i)$ defining the optimal prior-to-posterior update in \eqref{eq:optCovLinear},
though strongly data-informed, need not be relevant to the QoI.  For instance, some of the
$(w_i)$ could lie in the nullspace of the goal-oriented operator.  
Computing these
eigenvectors would be an unnecessary waste of computational resources.
Of
course, as the rank of the optimal prior-to-posterior update increases, the corresponding
approximation $\Gzysub$ will continue to improve until eventually $\Gzy =\Gzysub$.  In
the worst case scenario, however, $\Gzysub$ will be a good approximation of $\Gzy$ only as
we start computing eigenpairs of $(H,\Gpr^{-1})$ associated with the smallest nonzero
generalized eigenvalues.  This is clearly unacceptable as the overall complexity of the
approximation algorithm would not depend on the nature of the goal-oriented operator.
Therefore, the approximation \eqref{eq:approxCovQoIsub} cannot satisfy any reasonable
optimality statement in the spirit of Theorem \ref{thm:mainLinear} and calls for a proper
modification.  

\subsubsection{An optimal approximation}
\label{sec:optApproxCovQoI}
\hrevone{
The form of $\Gzysub$ in \eqref{eq:approxCovQoIsub} shows that 
the posterior covariance of the QoI can be written as a low-rank update of the prior
covariance. 
(Recall that the prior distribution of the QoI is Gaussian, 
$\Zb \sim \Gauss(0,\Gz)$ with $\Gz = \gop\,\Gpr \, \gop^\top$.)
This is  once again consistent with our intuition about the Bayesian update: the data
will only inform certain aspects of the QoI.
}
Thus a structure-exploiting approximation class
for $\Gzy$ is given by the set of positive definite matrices that can be written as
rank--$r$ negative-definite updates of $\Gz$:
	\begin{equation} \label{eq:classQoI}
		\Mc_r^{\Zb} = \{ \Gz - KK^\top \succ 0 : \rank(K)\le r \}.
	\end{equation}
\hrevone{Before introducing one of the main
results of this paper, we observe that $\Yb$ and $\Zb$ are related
by a linear model similar to  \eqref{eq:QoIstatistics}. 
The following lemma clarifies this relationship. (See
Appendix \ref{sec:proofs} for a proof.)
}
\smallskip

\begin{lemma} \label{lem:linearModelQoi}
A linear Gaussian model consistent with \eqref{eq:QoIstatistics}   is given by:  
\begin{equation} \label{eq:red_model}
\Yb = G \, \gopd \, \Zb + \errorTr,
\end{equation}
where $\gopd \coloneqq  \Gpr \gop^\top \Gz^{-1}$,
$\Zb \sim \Gauss(0,\Gz)$ 
  and $\errorTr \sim \Gauss( 0,  \Gamma_{\errorTr} ) $ are independent,
  and 
$\Gamma_{\errorTr} \coloneqq \Gobs + G (\Gpr - \Gpr \gop^\top \,\Gz^{-1} \,\gop \Gpr ) G^\top$. 
\end{lemma}
\smallskip

The results of \cite[Theorem 2.3]{spantini2014optimal} can be extended to the goal-oriented case 
by applying them to the reduced linear model in \eqref{eq:red_model}.
The following theorem defines the optimal approximation of $\Gzy$ and is one
of the main results of this paper. See Appendix \ref{sec:proofs} for
a proof.
\smallskip
\begin{theorem}[Optimal approximation of the posterior covariance of the QoI]
\label{approxCovUpdate-qoi}
Let $(\lambda_i , q_i)$ be the  
eigenpairs of:
\begin{equation} \label{pencilGoal1}
( G\, \Gpr \, \gop^\top \, \Gz^{-1} \,\gop \, \Gpr\, G^\top \, ,\, \Gy )
\end{equation}
with the ordering $\lambda_i \ge \lambda_{i+1}>0$ and normalization 
$q_i^\top \,G\, \Gpr \, \gop^\top \, \Gz^{-1} \,\gop \, \Gpr\, G^\top\, q_i = 1$, where 
$\Gy \coloneqq \Gobs + G\,\Gpr \,G^\top$ is the covariance matrix of the marginal
distribution of $\Yb$. Then, a minimizer $\Gzyopt$ of the 
Riemannian metric $\dr$ between $\Gzy$ and an element of ${\Mc}_r^{\Zb}$ is given by:
\begin{equation} \label{minimizer_theorem_covgoal}
 \Gzyopt=\Gz-KK^\top, \quad KK^\top=\sum_{i=1}^r \lambda_i \, \widehat{q}_i \widehat{q}_i^\top, \quad \widehat{q}_i \coloneqq \gop \,\Gpr\,G^\top q_i,
 \end{equation}
 where the corresponding minimum  distance is:
 \begin{equation} \label{eq:error_estimate_covgoal}
 \dr^2( \Gzy , \Gzyopt  )=  \frac{1}{2}	 \sum_{i>r}  {\rm ln}^2(\,1-\lambda_i\,).
 \end{equation}    
\end{theorem}
The optimal approximation in Theorem \ref{approxCovUpdate-qoi} yields the best 
approximation
for any given 
rank of the prior-to-posterior update and, most importantly, never requires 
the full posterior covariance of the parameters.
(This should be contrasted with 
\cite{lieberman2012}.)  
The directions $(q_i)$ that define the optimal update are  the leading eigenvectors
of  
$(G\, \Gpr \, \gop^\top \, \Gz^{-1} \,\gop \, \Gpr\, G^\top , \Gy )$ and  stem from a   careful balance of all the ingredients of the goal-oriented inverse problem:
the forward model, measurement noise, prior information, and ultimate goals.
Incorporating ultimate goals reduces the  intrinsic dimensionality  of the inverse
problem: for any fixed approximation error, the rank of the optimal update 
\eqref{minimizer_theorem_covgoal} can only be less than or equal to 
that of the suboptimal approximation introduced in
\eqref{eq:approxCovQoIsub}. 

\subsubsection{Computational remarks}
\label{sec:comp_rem}
If square roots of $\Gpr$ and $\Gobs$ are available,
such that
$\Gpr = \Gprs \,\Gprs^\top$ and $\Gobs = \Gobss \,\Gobss^\top$, then
we can rewrite  the pencil  \eqref{pencilGoal1} in a more
concise form as follows.

\begin{corollary} \label{cor:equivPencils}
Let $\widehat{G}\coloneqq \Gobss^{-1}\,G\,\Gprs$, and 	
let $\Pi$  be an orthogonal projector onto the
range of $\Gprs^\top\,\gop^\top$. Then the eigenvalues of
\begin{equation} \label{pencilGoal1equiv}
   (\,\widehat{G}\,\Pi\,\widehat{G}^\top\,, 
   \,I+\widehat{G}\,\widehat{G}^\top\,)	
\end{equation}
are the same as those of \eqref{pencilGoal1}, and the eigenvectors of
\eqref{pencilGoal1equiv} can be mapped to the eigenvectors of \eqref{pencilGoal1}
with the transformation $w \mapsto \Gobss^{-\top} w$. 
\end{corollary}
\smallskip

The proof of the corollary is straightforward once we note that  
$\Pi$ can be written as
$\Pi = \Gprs^\top\,\gop^\top\, (  \gop  \Gprs \Gprs^\top  \gop^\top )^{-1}  \,\gop\,\Gprs =  \Gprs^\top\,\gop^\top\,\Gz^{-1}\,\gop\,\Gprs$.
Moreover, the action of the projector $\Pi$ on a vector
$v$ can be computed efficiently since 
$\Pi(v) \coloneqq \Gprs^\top\,\gop^\top x_{\rm ls}$, where
$x_{\rm ls}$ is the least squares solution of the overdetermined
linear system $\Gprs^\top\,\gop^\top\,x_{\rm ls} = v$.
There is a variety of techniques for the solution of large-scale
matrix-free least squares problems 
(e.g., \cite{paige1982lsqr,fong2011lsmr,choi2006iterative,hestenes1952methods,meng2014lsrn}).

We now focus our computational remarks on the analysis of the
pencil in \eqref{pencilGoal1equiv}, which is well suited for practical implementations of
the approximation.  To simplify notation, let us rewrite \eqref{pencilGoal1equiv} as
$(A,B)$, where $A\coloneqq \widehat{G}\,\Pi\,\widehat{G}^\top$ and
$B \coloneqq I+\widehat{G}\,\widehat{G}^\top$.

Finding the leading generalized eigenpairs of \eqref{pencilGoal1equiv} requires the solution
of a Hermitian generalized eigenvalue problem \cite{bai2000templates}.
Unfortunately, it is not easy to reduce \eqref{pencilGoal1equiv} 
to a standard eigenvalue problem,\footnote{%
In contrast, this is often  possible in the non-goal-oriented case when dealing with the
pencil $(H,\Gpr^{-1})$, as the action of a square root of $\Gpr^{-1}$, or of its inverse, is available in many cases of interest (e.g., \cite{lindgren2011explicit}).
} 
as doing so would require the action of a square root of $B$ or of $B^{-1}$. Nevertheless,
there are a plethora of matrix-free algorithms for large-scale generalized eigenvalue
problems: generalized Lanczos iteration \cite[Section~5.5]{bai2000templates}, randomized
SVD--type methods \cite{saibaba2013randomized}, manifold optimization algorithms
\cite{absil2006truncated,absil2005adaptive,baker2006implicit}, the trace minimization
algorithm \cite{sameh1982trace,klinvex2013parallel}, and the inverse--free preconditioned
Krylov subspace method \cite{golub2002inverse}, to name a few. These algorithms require
the iterative solution of linear systems associated with $B$, in some cases to low
accuracy \cite{sameh1982trace,golub2002inverse}.
\hblue{ 
Applying $B$ to a vector requires the evaluation of the forward model, which may 
or may not be expensive 
(e.g., consider PDE based models \cite{flath2011fast} versus image deblurring problems
\cite{kaipio2007statistical}). 
In practice, for a \emph{fixed} dimension of the desired
eigenspace, algorithms for characterizing the eigenpairs of $(A,B)$ 
lead to more expensive
computations than those for the pencil $(H,\Gpr^{-1})$ used in the
na\"{i}ve approximation of Section \ref{sec:naiveApprox}.
However, the key point is that the 
optimal approximation of Theorem \ref{approxCovUpdate-qoi} 
requires the characterization of lower dimensional eigenspaces
for a given accuracy.
}

 Moreover, if we solve the generalized eigenvalue problem
using a block Lanczos iteration or a randomized method, then we can
also exploit block Krylov methods to solve the associated linear
systems---comprised of $B$ and multiple right-hand
sides---simultaneously \cite{o1980block,saad2003iterative}.
In particular, the convergence of Krylov methods for solving linear systems of the
form $Bx = b$, such as the conjugate gradient algorithm, depends
not only on the spectrum of $B$ but also on the right-hand side $b$
\cite{liesen2004convergence,axelsson2000sublinear}.  This dependence is
especially important when the right-hand side has some structure
and is not entirely random: in our case, $b$ lies in the range of the
possibly low-rank operator $A$.  For instance, if $b$ is {\it mostly}
contained in a low-dimensional invariant subspace of $B$ (whether
associated with small or large eigenvalues), then the Krylov solver
will likely converge to an accurate solution in few steps.
Conversely, it should be noted that if the range of the operator $\Pi$ in
\eqref{pencilGoal1equiv}---essentially the subspace of the parameter space that is
relevant to the QoI---has non-negligible components along \textit{every} data-informed
parameter direction (corresponding to the leading eigenspace of
$(H,\Gpr^{-1})$), then it would be difficult to obtain an accurate approximation of
$\Gzy$ without exploring the full data-informed subspace.  %
\hblue{ 
Even though the na\"{i}ve approximation
is suboptimal,
it may be interesting from a practical standpoint to assess
its performance, since it is cheaper
to compute for a given approximation rank.}
The
following lemma provides useful guidelines in this direction. See Appendix
\ref{sec:proofs} for a proof of this result.  \smallskip

\begin{lemma}[Relationship between approximations]
\label{lem:comparisonApprox}
Let $\Gzyopt, \ \Gzysub \in {\Mc}_r^{\Zb}$ be, respectively, the optimal and suboptimal
approximations of $\Gzy$ introduced in \eqref{minimizer_theorem_covgoal} and
\eqref{eq:approxCovQoIsub}.  Moreover, let $\Gposh \in {\Mc}_r$ be the optimal
approximation of $\Gpos$ defined in \eqref{eq:optCovLinear}. Then
\begin{equation} \label{eq:comparisonDistances}
	 \dr(\, \Gzy \, , \,\Gzyopt  \, ) \le \dr( \, \Gzy \, , \, \Gzysub \,) \le
	 \dr(\, \Gpos \, , \, \Gposh \,).
\end{equation}
\end{lemma}
\smallskip
Lemma \ref{lem:comparisonApprox} has several interesting consequences.  First of all, notice that
it is possible to bound the accuracy of the na\"{i}ve approximation,
$\Gzysub=\gop\,\Gposh\,\gop^\top$, using $\dr(\, \Gpos \, , \, \Gposh \,)$.  The latter
distance can easily be bounded as a function of the generalized eigenvalues of
$(H,\Gpr^{-1})$, as shown in \eqref{eq:distRlinearpaper}.  These are precisely the
eigenvalues computed by the na\"{i}ve approximation.  Thus, if the eigenvalues of
$(H,\Gpr^{-1})$ decay sharply
or, equivalently, if the distance $\dr(\, \Gpos \, , \, \Gposh \,)$ can be made small with
only a low-dimensional (small $r$) prior-to-posterior update, then Lemma
\ref{lem:comparisonApprox} says that the na\"{i}ve approximation, albeit suboptimal, can
yield a remarkably efficient approximation of $\Gzy$---with strong accuracy guarantees in
terms of $\dr(\, \Gpos \, , \, \Gposh )$. Intuitively, if $\Gzysub$ already accounts for
most of the data-informed directions in the parameter space, then there is no major
loss of accuracy in neglecting further directions of the prior-to-posterior update, even
if these directions are relevant to the QoI. In this situation, these additional directions
would provide very limited information relative to the prior and can be safely
neglected.

On the other hand, if the eigenvalues of $(H,\Gpr^{-1})$ do not decay as quickly, then the
bound provided in \eqref{eq:comparisonDistances} becomes useless.  This is not to say that
the na\"{i}ve approximation will necessarily perform poorly.  (It is possible that the QoI
depends \textit{only} on a few of the leading data-informed directions, such that the
na\"{i}ve approximation performs well even for a low rank prior-to-posterior update.)  But
we cannot quantify the accuracy of the na\"{i}ve approximation unless we directly compute
$\dr(\Gzy,\Gzysub)$, which in turn requires the solution of an expensive generalized
eigenvalue problem for each rank of the update.  This is not feasible in practice. In such
situations, we should resort to the optimal approximation introduced in Theorem
\ref{approxCovUpdate-qoi}, which offers a useful error bound as well as a concrete
possibility for both computational and storage savings.

\subsubsection{
Properties of the optimal covariance approximation}

An important consequence of the optimal approximation of $\Gzy$ with respect to
the metric $\dr$ is optimality \textit{in distribution} whenever the posterior mean of the QoI
is known. 
It follows from \cite[Lemma~2.2]{spantini2014optimal} that
the minimizer of the Hellinger distance (or the Kullback--Leibler divergence) between the
Gaussian posterior measure of the QoI, $\mzy \coloneqq \Gauss(\muzy , \Gzy)$, and the approximation 
$\Gauss( \muzy ,  \Gamma  )$ for a 
matrix $ \Gamma \in\Mc_r^{\Zb}$, is given by the optimal approximation 
\eqref{minimizer_theorem_covgoal} defined
in Theorem \ref{approxCovUpdate-qoi}.
In particular, let $\mzya \coloneqq \Gauss( \muzy ,  \Gzyopt )$ be the measure that
optimally approximates $\mzy$, where $\Gzyopt$ is defined in \eqref{minimizer_theorem_covgoal}.
Then it is easy to show that the Hellinger distance between 
$\mzy$ and the optimal approximation $\mzya$ is given by:
 \begin{equation} \label{eq:HellQoI}
  d_{\rm Hell}(\mzy,\mzya) = \sqrt{  1 - \prod_{i>r} 
  2^{1/2}  \frac{ (1-\lambda_i)^{1/4} }
   { (2 -   \lambda_i   )^{1/2} } }
 \end{equation}
where $(\lambda_i)$ are the generalized eigenvalues defined in  
Theorem \ref{approxCovUpdate-qoi} (e.g., \cite[Appendix A]{spantini2014optimal}
or \cite{pardo2005statistical}).

The Hellinger distance can be used to bound the error of expectations of
functions of interest with respect to approximate measures \cite{dashti2013bayesian}.  
That is, suppose that we are
interested in the posterior expectation $\Ex_{\mzy}[g]$ of some measurable function
$g:\re^p\ra\re$, with certain bounded moments with respect to the prior measure
$\nuz\coloneqq \Gauss(0,\Gz)$, and suppose further that we can only evaluate integrals
with respect to the approximate measure $\mzya$.  Then the error resulting from computing
$\Ex_{\mzya}[g]$, as opposed to $\Ex_{\mzy}[g]$, for a fixed realization of the data
$\Yb$, can be bounded in terms of the Hellinger distance between the
two \hrevone{Gaussian} measures using
the following lemma, which
follows easily from
\cite[Lemma 7.14]{dashti2013bayesian}.
(See Appendix \ref{sec:proofs} for a proof.)  \smallskip

\begin{lemma}[Convergence in expectation]
\label{lem:convergHell}
Let $g:\re^p \ra \re$ be a measurable function with $\beta>2$ bounded moments
with respect to the prior measure, i.e., $\Ex_{\nuz}[\,|g|^\beta\,]<\infty$. 
Then:
\begin{equation} \label{eq:boundHell}
\left \vert  \Ex_{\mzy}[g] - \Ex_{\mzya}[g]  \right \vert  \le 
\mathcal{C}(\Yb,g)\,d_{\rm Hell}(\mzy,\mzya), 
\end{equation}
where $\mathcal{C}(\Yb,g) \coloneqq 2\,\sqrt{2}\, \frac{|\Gz|^{1/4}}{|\Gzy|^{1/4}}  \, 
\exp \left( \frac{1}{2(\beta-2)} \,\| \muzy \|_{\Gz^{-1}}^2 \right)\,
\Ex_{\nuz}[\,|g|^\beta\,]^{1/\beta}$ and where $|A|$
denotes the determinant of the matrix $A$. 
\end{lemma}
\smallskip

Notice that the constant $\mathcal{C}(\Yb,g)$ in \eqref{eq:boundHell} is independent of the
approximating measure $\mzya$.
Convergence of the approximation in Hellinger distance thus
implies convergence of the expectation $\Ex_{\mzya}[g]$ to $\Ex_{\mzy}[g]$.  \medskip

It is interesting to note that the optimal approximation of the posterior covariance matrix of the QoI  in Theorem \ref{approxCovUpdate-qoi}
is always associated with a corresponding approximation of the posterior covariance of
the parameters, $\Gpos$.
The following result clarifies the nature of this approximation.
It is 
proved in Appendix \ref{sec:proofs}.
\smallskip

\begin{lemma}[Goal-oriented approximation of 
$\Gpos$]
\label{cor:optGoalApproxCov}
Let $\Gzyopt$ be the minimizer of the metric $\dr$ between $\Gzy$ and an element of ${\Mc}_r^{\Zb}$
as given by \eqref{minimizer_theorem_covgoal}. Then $\Gzyopt$ can be written as
	\begin{equation} \label{eq:goalOrientedGpos}
		\Gzyopt = \gop\, \Gposh^* \, \gop^\top, \qquad 
		\Gposh^* = \Gpr - \sum_{i=1}^r \, \lambda_i \, 
		\widetilde{q}_i \, \widetilde{q}_i^\top, \qquad
		\widetilde{q}_i  \coloneqq \Gprs\,\Pi \,\Gprs^\top \,G^\top q_i,
	\end{equation}
where the vectors $(q_i)$ are defined in Theorem \ref{approxCovUpdate-qoi}, 
$\Gprs$ is a square root of the prior covariance matrix such that $\Gpr = \Gprs\,\Gprs^\top$,
$\Pi$ is the orthogonal projector onto the
range of $\Gprs^\top\,\gop^\top$,
while 
$\Gposh^*$ satisfies
\begin{eqnarray}  \label{eq:goalOrientedForstner}
	\Gposh^*\,\, &\in & {\rm arg}\, {\min_\Gamma}\,\dr(\, \Gzy \, ,\, \gop \,\Gamma\,\gop^{\top} )\\
	                &{\rm s.t.} & \,\, \Gamma \in \Mc_r \coloneqq 
	                \{ \Gamma = \Gpr - K\,K^\top \succ 0, \,\rank(K)\le r \}. \nonumber
\end{eqnarray}
\end{lemma}
\smallskip
The matrix $\Gposh^*$ is an optimal {\it goal-oriented} approximation of $\Gpos$. 
This notion of optimality is quite different from that
 in Theorem  \ref{thm:mainLinear}.
The prior-to-posterior update directions, $(\widetilde{q}_i)$ in
\eqref{eq:goalOrientedGpos},
have
the  intuitive interpretation of directions, in the parameter space, that are most informed
by the data, relative to the prior, \textit{and} that are relevant to the QoI.
In particular, it is easy to see that the  $(\widetilde{q}_i)$ 
are orthogonal to the nullspace of the goal-oriented operator with respect to the inner product induced by the prior precision, i.e., 
\begin{equation}
	h^\top \Gpr^{-1}\, \widetilde{q}_i = (\gop\,h)^\top \Gz^{-1} \gop \Gpr G^\top q_i = 0, 
	\qquad \forall h \in {\rm Null}(\gop).
\end{equation}
Note that even if $\Gzyopt = \gop \,\Gposh^*\, \gop^\top$ 
is a good approximation of $\Gzy$,  $\Gposh^*$ need not be	 
a good approximation of $\Gpos$.

Now we introduce a particularly simple factorization of the optimal approximation
$\Gzyopt$ from Theorem \ref{approxCovUpdate-qoi}, as $\Gzyopt = \Gzyopts \,\Gzyopts^\top$
for some matrix $\Gzyopts$.  We can think of $\Gzyopts$ as a square root of
$\Gzyopt$, even though $\Gzyopts$ need not be a square matrix.  Obtaining the action of a
square root of $\Gzyopt$ on a vector is an essential task if our goal is to sample the
distribution $\Gauss( \muzy , \Gzyopt )$ in truly high-dimensional problems.
The key requirement is that $\Gzyopts$ be easy to compute once we have the optimal
approximation $\Gzyopt$.  We have deferred the discussion of this topic until now in order
to exploit the results of Lemma \ref{cor:optGoalApproxCov} to obtain an explicit
characterization of $\Gzyopts$.  The proof of the following lemma can be found in Appendix
\ref{sec:proofs}.  \smallskip

\begin{lemma} \label{cor:sqrtGyz}
Let $(\lambda_i , q_i)$, $\Gprs$, and
$\Pi$ be defined as in 
Lemma \ref{cor:optGoalApproxCov}.
Then, a non-symmetric square root, $\Gzyopts$, of $\Gzyopt$, 
such  that $\Gzyopt = \Gzyopts \,\Gzyopts^\top$, is given by
\begin{equation} \label{eq:formSqrtGyz}
	\Gzyopts =  \gop\,\Gprs\,\left(\,\sum_{i=1}^r\,
	(\sqrt{1-\lambda_i}-1)\,\bar{q}_i\,\bar{q}_i^\top + I\, \right), \qquad
	\bar{q}_i \coloneqq \Pi \,\Gprs^\top \,G^\top q_i,
\end{equation}
where $I$ is the identity matrix.
\end{lemma}
\smallskip

The virtue of this result is that it does not require an 
invertible square root of $\Gz=\gop\,\Gpr\,\gop^\top$ as one would 
expect from \cite[Remark 2]{spantini2014optimal}.
Note that it is easy to apply $\Gzyopts$ to a vector, which
allows efficient sampling from $\Gauss( \muzy , \Gzyopt )$.
An interesting feature of  $\Gzyopts\in\re^{p\times n}$ is that it
is a nonsquare matrix with $p<n$. 
This is certainly not an issue as long as $\Gzyopts \,\Gzyopts^\top = \Gzyopt$.
Notice also that \eqref{eq:formSqrtGyz} contains a square root of the
prior covariance matrix.
The action of this matrix is usually available in large-scale applications (e.g., \cite{wood1994simulation,dietrich1997fast,lindgren2011explicit,stuart2010inverse,yue2010nonstationary}).
However, if the action of a square root of $\Gpr$ is truly unavailable, then
one can still sample from $\Gauss( \muzy , \Gzyopt )$ by resorting to the 
action of the matrix $\Gzyopt$ alone (e.g., 
\cite{chow2014preconditioned,fox2014convergence,parker2012sampling,schneider2003krylov}).
It is  straightforward to apply 
$\Gzyopt$ to a vector (see \eqref{minimizer_theorem_covgoal}).

\subsection{Approximation of the posterior mean of the QoI}
\label{sec:approx_mean_QoI}
We conclude this theory section by introducing an optimal approximation of the posterior
mean of the QoI. The cost of computing
\begin{equation}
  \muzy\coloneqq \gop \,\mupos = \gop\,\Gpos\, G^\top \Gobs^{-1} \,\Yb
\end{equation}
for a single realization of the data is usually dominated by the cost of solving a single
linear system associated with $\Gpos^{-1}$ to determine $\mupos$.  This task can be
efficiently tackled with state-of-the-art matrix-free iterative solvers for symmetric
linear systems (e.g., \cite{bai2000templates,hestenes1952methods,akccelik2006parallel})
even for million-dimensional parameter spaces \cite{bui2012extreme}.  If, however, one is
interested in the fast computation of $\muzy$ for multiple realizations of the data (e.g.,
in the context of online inference), then the situation is quite different
\cite{chung2014efficient,friedland2007generalized,markovsky2008structured,hua1998generalized, chung2014optimal,chung2013computing}.  Solving a linear system to
compute $\muzy$ each time a new measurement is available might be infeasible in practical
applications.  If the dimension of the QoI is small, say $p=O(1)$, then there is an easy
solution to this problem.  One can just precompute the matrix
$M\coloneqq \gop\,\Gpos\, G^\top \Gobs^{-1}$ in an offline stage and then compute the
posterior mean of the QoI as $\muzy = M\,\Yb$ each time a new realization of the data
becomes available.  Yet the computational efficiency of this procedure breaks down as the
dimension of the QoI increases---for instance, if the QoI is a finite-dimensional
approximation of some underlying function. In this case, the matrix $M$ would be large and
dense, and storing it could be quite inefficient.  Moreover, performing a dense
matrix-vector product to compute $\muzy = M\,\Yb$ might become more expensive than solving,
a single linear system associated with $\Gpos^{-1}$.

Our goal is thus to characterize computationally efficient and statistically optimal
approximations of $\muzy$.
In particular, we seek an approximation of $\muzy$ as a low-rank linear function of the
data,
i.e., $\muzy \approx \muzyapp \coloneqq A\,\Yb$ for some low-rank matrix $A$.  With such
an $A$, computing $\muzyapp$ for each new realization of the data would be computationally
efficient.  
We define optimality of the approximation with respect to the Bayes risk for squared-error
loss weighted by the posterior precision matrix of the QoI,
i.e.,
	\begin{equation} \label{eq:formBayesrisk}
		\Bc(A)\coloneqq\mathbb{E} \left [  \Vert  A \, \Yb - \Zb
                  \Vert_{\Gzy^{-1}}^2 \right ], 
	\end{equation}
where $\Bc(A)$ denotes the Bayes risk associated with the matrix $A$, 
and where the expectation is taken over the joint distribution of $\Zb$ and $\Yb$.
\hrevone{  
(The minimizer of the Bayes risk for squared error loss over all linear
functions of the data is precisely the posterior mean of the QoI.)
}
Minimizing the Bayes risk \eqref{eq:formBayesrisk} is equivalent to minimizing
\begin{equation} \label{eq:approxMeanQoI}
	\Ex \left [  \| \muzy - \muzyapp  \|_{\Gzy^{-1}}^2  \right ]
\end{equation}
over all approximations of the posterior mean of the form $\muzyapp = A\,\Yb$
for some low-rank matrix $A$.
The Mahalanobis distance in \eqref{eq:approxMeanQoI} is precisely a Riemannian metric 
of the form described in Section \ref{s:metrics} and thus it is a natural way 
to assess the quality of a posterior mean approximation.
In particular, the weighted norm in \eqref{eq:approxMeanQoI} penalizes errors in 
the approximation of $\muzy$ more strongly in directions of lower posterior variance.
As a  result,  the approximation of $\muzy$ is more likely to fall within the
bulk of the posterior density of the QoI.
Notice that \eqref{eq:approxMeanQoI} is
an average of the squared Riemannian distance between
$\muzy$ and its approximation $\muzyapp$ over
the distribution of the data $\Yb$.

The following theorem characterizes the optimal approximation of $\muzy$.
See Appendix \ref{sec:proofs} for  a proof.
\smallskip

\begin{theorem}[Optimal approximation of $\muzy$] 
\label{thm:mean_approx_lowrank}
Let $(\lambda_i , q_i, \widehat{q}_i)$ be defined as in Theorem \ref{approxCovUpdate-qoi} 
and consider the minimization of the following Bayes risk over the set of low-rank matrices:	
\begin{equation} \label{eq:optimMean}
	\min_A \ \mathbb{E} \left [ \Vert  A \, \Yb - \Zb  \Vert_{\Gzy^{-1}}^2 \right  ], 
	\qquad  {\rm s.t.\quad} {\rm rank}(A)\le  r \, .
\end{equation}
Then a minimizer of \eqref{eq:optimMean} is given by:
 	\begin{equation} \label{minimizerMeanLowRank}
     	A^\ast =  \sum_{i= 1}^r  \lambda_i \, \widehat{q}_i \, q_i^\top,
 	\end{equation}
with minimum Bayes risk:
	\begin{equation} \label{bayesRiskOptimum}
			\Bc(A^*)= 
			\mathbb{E} \left [  \Vert \,  A^* \, \Yb - \Zb  \, \Vert_{\Gzy^{-1}}^2
                          \right  ]  = 
			\sum_{i>r} \, \frac{\lambda_i}{1-\lambda_i}  +
                        n,
	\end{equation}	  	
where $n$ is the dimension of the parameter space.	
\end{theorem}
\smallskip

Note that \eqref{minimizerMeanLowRank} can be computed ``for free'' from the optimal
approximation of $\Gzy$ introduced in Theorem \ref{approxCovUpdate-qoi}.  Also, the
optimal approximations of both the posterior mean and the posterior covariance of the QoI
become quite accurate as soon as we start including generalized eigenvalues
$\lambda \ll 1$ in the corresponding approximations (see minimum loss
\eqref{eq:error_estimate_covgoal} and Bayes risk \eqref{bayesRiskOptimum}).

\section{Proof-of-concept example}
\label{s:simpleExample}
\begin{figure}[!htp]
\begin{center}
\includegraphics[width=0.44\textwidth]{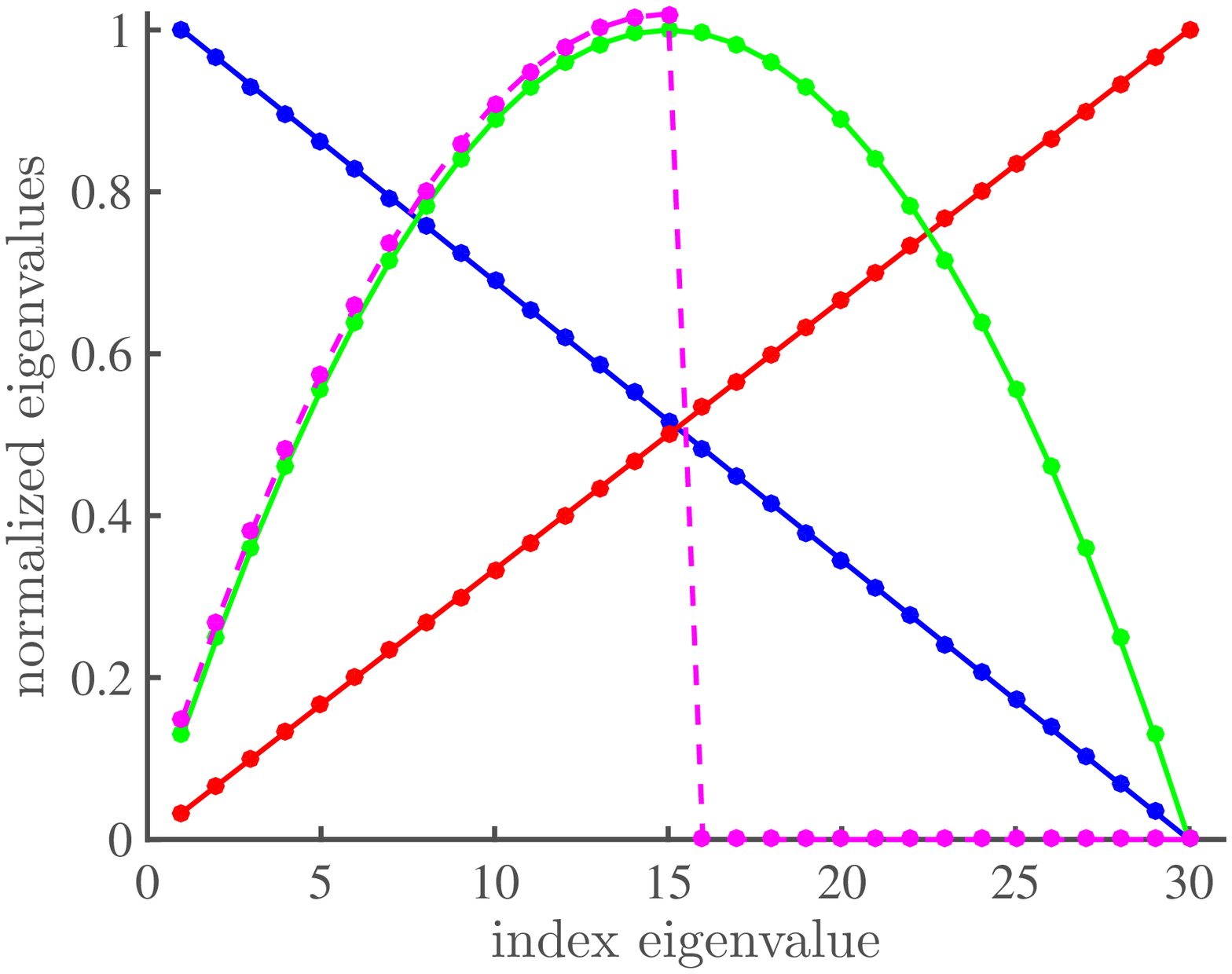}
\caption{
\hrevone{ 
Normalized eigenvalues defined in the
proof-of-concept example of Section \ref{s:simpleExample}: 
in blue we show $(h_i/h_{\rm max})$, in red $(\mu_i/\mu_{\rm max})$,
in green $(\delta_i^2/\delta_{\rm max}^2)$, 
and in magenta $(\hat{\lambda}_i/\hat{\lambda}_{\rm max})$, for
$i=1,\ldots,n$ and $n=30$.
For any finite collection of eigenvalues $(\sigma_i)$, we define 
$\sigma_{\rm max}$ to be the maximum value over that collection.	
Since for $i \le p$ and $p=15$, we have $\hat{\lambda}_i=\delta_i^2$
in this example, we shifted the magenta curve slightly upwards to
distinguish it from the green one.
}	
}
\label{fig:proof-concept} 
\end{center}
\end{figure}

\hrevone{ 
Before investigating the numerical performance of our goal-oriented approximations, we
illustrate the theory with a simple proof-of-concept example.
We consider an identity forward model $G=I$, a diagonal
observational noise precision $\Gobs^{-1} = {\rm diag}(h_1,\ldots,h_n)$, and
a diagonal prior covariance $\Gpr = {\rm diag}(\mu_1,\ldots,\mu_n)$, with $h_i= n - i$ and $\mu_i = i$ for 
$i=1,\ldots,n$. We may think of this problem as denoising a signal $\Xb$ \cite{spantini2014optimal}.
Figure \ref{fig:proof-concept} shows the normalized eigenvalues of $\Gobs^{-1}$ and $\Gpr$
in blue and
red, respectively, for the case $n=30$.
The eigenvectors of 
both matrices correspond to the canonical vectors in $\re^n$, i.e., 
$\eb_1,\ldots,\eb_n$.
In this case,  the data are most informative---in absolute terms---along
directions $\eb_i$ with $i \ll n$, since the observational noise
precision $h_i$ is a decreasing function of $i$.
On the other hand, the prior variance is large along $\eb_i$ when $i \gg 1$, since
$\mu_i$ is an increasing function of $i$. Thus
the prior is more constraining where the data are more
informative.
The eigenpairs $(\delta_i^2,w_i)$ of the pencil $(H,\Gpr^{-1})$, defined
in Theorem \ref{thm:mainLinear}, are given by
$\delta_i^2 = h_i \cdot \mu_i = (n-i)\cdot i$ and $w_i \propto \eb_i$ for $i=1,\ldots,n$.
(These $\delta_i^2$ are not sorted in decreasing order; for
simplicity, we retain the same index $i$ as in the problem definition.)
From the relative magnitudes of $(\delta_i^2)$---illustrated by the
green parabola in Figure \ref{fig:proof-concept}---we can identify the
parameter directions that are most informed by the data {\it relative}
to the prior: they correspond to $\eb_i$ with $i$ around $n/2$ (the
middle of the spectrum).  These directions define the optimal
prior-to-posterior update of Theorem \ref{thm:mainLinear}.
Modes $\eb_i$ with $i \ll n/2$ are strongly informed by the data in an
absolute sense, but not {\it relative} to the prior; thus their
overall importance is limited. In the same way, modes $\eb_i$ with
$i \gg n/2$ are unimportant to the update since the posterior variance
along these directions is roughly equal to the prior variance
$(\delta_i^2\ll 1)$, even though both variances are relatively large
\cite{spantini2014optimal}.}

\hrevone{ 
Now let the goal-oriented operator $\gop:\re^n \ra \re^p$ be defined
as follows: $\gop \xb=(x_1 , \ldots, x_p)$ for
$\xb=(x_1,\ldots,x_n)$ and $p=n/2$.
Simple algebra shows that the
goal-oriented eigenpairs $(\lambda_i , q_i)$ of Theorem
\ref{approxCovUpdate-qoi} are given by $q_i\propto\eb_i$ for
$i=1, \ldots, n$, and $\lambda_i = 1/(1+1/(h_i \mu_i))$ 
for $i\le p$ and $\lambda_i = 0$ for $i>p$.
So that the eigenvalues $\delta^2_i$ and $\lambda_i$ are comparable
in terms of their associated covariance approximation errors---see
\eqref{eq:distRlinearpaper} and \eqref{eq:error_estimate_covgoal}---we
plot a nonlinear function of each $\lambda_i$ in Figure
\ref{fig:proof-concept}, namely $\hat{\lambda}_i=f(\lambda_i)$ for
$f(x)=1/(1-x)$.
(Since $f$ is strictly increasing on $[0,1)$, the relative importance
of the $(\hat{\lambda}_i)$ is the same as that of the original
$(\lambda_i)$.)
The introduction of a goal-oriented operator reveals directions that
can be strongly informed by the data, relative to the prior, but that
are irrelevant to the QoI.  These modes correspond to $(\eb_i)$ for
$i>p$, and can be safely neglected when computing the Bayesian update
relevant to the QoI.}  

\hrevone{
Of course, in the general case of non-diagonal operators $(G,\Gobs,\Gpr)$, the directions $(w_i)$ and $(q_i)$ need
not coincide. The following numerical example will illustrate this general situation.
}

\section{Numerical examples}
\label{s:examples}

Now we numerically illustrate the performance of our approximations using a goal-oriented inverse problem in heat transfer.
In particular, we study the cooling of a CPU by means of a heat sink. Our goal is to infer
the (spatially inhomogeneous) temperature of the CPU from noisy pointwise observations of
temperature on the heat sink. Figure \ref{fig:heat_configuartion} shows the problem setup:
the three different layers of material correspond, respectively, to the CPU ($\Dom_{1}$),
a thin silicone layer that connects the CPU to the heat sink ($\Dom_{2}$), and an aluminum fin
($\Dom_{3}$).  We denote by $\Dom$ the union of these domains.  Each $\Dom_{i}$ represents
a two-dimensional cross section of material of constant width $W$ along the horizontal
direction and a height $L_{i}$.  We assume that no heat transfer happens along
the third dimension; this is a common engineering approximation
\cite{bergman2011fundamentals}.  Each material has a constant density $\rho_{i}$, a
constant specific heat $c_{i}$ and a constant thermal conductivity $k_{i}$.
The corresponding thermal diffusivities $\alpha_i = k_i / \rho_i c_i$ are shown in the
table at the right of Figure \ref{fig:heat_configuartion}. 
The time-dependent temperature field in each domain is $\Temp^{(i)}: 
\Dom_{i} \times T \rightarrow\mathbb{R}$, where $T =  (0,t_{\text{end}}]$, for  $i=1,2,3$. Jointly,
these temperature fields are simply $\Temp: \Dom \times T\rightarrow \mathbb{R}$. 

\subsection{Forward, observational and prior models}
The time evolution of each temperature field $\Temp^{(i)}$ is described by a linear
time-dependent PDE of the form
\begin{equation} \label{eq:heat_eq}
  \rho_{i} \, c_{i} \, \partial_t \, \Temp^{(i)}=\nabla \cdot (k_{i}\nabla \Temp^{(i)}), \qquad  i=1,\ldots,3,
\end{equation}
where $\partial_t$ denotes partial differentiation with respect to time.  We assume no
volumetric heat production and use Fourier's law for the heat flux
\cite{guglielmini1990elementi}.  Equations \ref{eq:heat_eq} should be complemented with
appropriate boundary and initial conditions to define a well-posed forward problem. We use
the independent variables $s_{1}$ and $s_{2}$ to denote, respectively, the horizontal and
vertical directions and let $\sbo =(s_{1},s_{2})$.  The point $\sbo=(0,0)$ corresponds to
the lower left corner of $\Dom$.  At the lower boundary of $\Dom_{1}$ we impose a space-
and time-dependent heat flux: $k_{1} \,\partial_{\vec{n}}\, \Temp^{(1)}=q(\sbo,t) $ for
$\sbo\in\Dom_{1,\text{bottom}}$, where $\vec{n}$ refers to the outward pointing normal and
$q$ is a given nonconstant scalar function in $\sbo$. 
At the interface between domains $\Dom_{i}$ and $\Dom_{i+1}$ we assume heat transfer by
conduction with no thermal contact resistance:
$k_{i} \, \partial_{\vec{n}} \,\Temp^{(i)} = k_{i+1} \,
\partial_{\vec{n}} \, \Temp^{(i+1)}$ and
$\Temp^{(i)} = \Temp^{(i+1)}$
for $\sbo\in\text{interface}(\Dom_{i},\Dom_{i+1})$ and $i=1,2$.
At the top, left, and right boundaries of $\Dom_{3}$, we assume
heat transfer by convection with a fluid at constant temperature $\Temp_{\infty}$:
$-k_{3} \,\partial_{\vec{n}} \, \Temp^{(3)} =h_{c}(\Temp^{(3)}-\Temp_{\infty})$ for
$\sbo\in\Dom_{3,\text{top}}\cup\Dom_{3,\text{left}}\cup\Dom_{3,\text{right}}$,
where $h_{c}$ is a convective heat transfer coefficient. 
Finally, we impose adiabatic conditions (no heat exchange) on the left and right boundaries
of $\Dom_{1}$ and $\Dom_{2}$: 
$\,\partial_{\vec{n}} \, \Temp^{(i)} =0 $ for
$\sbo \in\Dom_{i,\text{left}}\cup\Dom_{i,\text{right}}$ and $i=1,2$.
The initial conditions are not specified here as they are the objective of the forthcoming
inverse problem.

We consider a finite element spatial approximation of the weak form of 
\eqref{eq:heat_eq} by means of linear
elements on simplices \cite{quarteroni2008numerical}.
We denote by $\Tempb_h(t)\in \re^n$ the collection of temperature values at the finite
element nodes on $\Dom$ at time $t\in T$.
The function
 $\Tempbh$ satisfies a system of ODEs of the form
$M \, \partial_{t}\Tempbh(t)+
	A \, \Tempbh(t)=\fb(t)$, with $t\in T$, 
for a suitable mass matrix $M$, stiffness matrix $A$, known time-dependent forcing term $\fb$ and initial conditions 
$\Tempbhz \coloneqq \Tempbh(t=0)$.
The initial conditions $\Tempbhz$ are unknown and must be estimated from local
measurements of the temperature field $\Temp$ at different locations in space and time.
The locations of the sensors $s^{1},\ldots,s^{N}$ are shown as black dots in Figure
\ref{fig:heat_configuartion}. Observations are collected every $\Delta t$ time units for
$t\in T$.  The first observation happens at time $t=\Delta t$ and we assume that there are
$M$ observation times in total. We denote measurements at time $t_{i}=i\Delta t$ as
$\widehat{\Yb}_{i}=\left[\Temp(s^{1},i\Delta t),\ldots,\Temp(s^{N},i\Delta t)\right]$.  We
concatenate the observations into a vector
$\widehat{\Yb}=(\widehat{\Yb}_{1},\ldots,\widehat{\Yb}_{M})\in\mathbb{R}^{d}$.  The actual
observations are corrupted with additive Gaussian noise:
$\Yb=\widehat{\Yb}+\error$, where $\error\sim\mathcal{N}(0,\sigma_{\text{obs}}^{2}\, I)$
and $I$ is the identity matrix.  Notice that $\widehat{\Yb}$ is an affine function of
$\Tempbhz$.  This relationship can be made linear by a suitable redefinition of the data
vector.  Thus, we are lead to a linear Gaussian inverse problem in standard form,
$\Yb =G\,\Tempbhz+\error$, where $G$ defines the forward operator,
$\Tempbhz \mapsto \widehat{\Yb}$, that can be evaluated implicitly by solving a heat
equation with no forcing term and initial conditions $\Tempbhz$ for a time interval
necessary to collect the corresponding observations $\widehat{\Yb}$.

We define a zero-mean Gaussian prior distribution\footnote{%
  There is no loss of generality in assuming zero prior mean. If we are given a
  statistical model of the form $\Yb =G\,\Tempbhz+\error$, where
  $\Tempbhz \sim \Gauss( \mupr , \Gpr )$ has a nonzero prior mean, then we can trivially
  rewrite this model as
  $\widehat{\Yb}\coloneqq \Yb - G\mupr =G\,(\Tempbhz-\mupr) + \error$ for a modified data
  vector $\widehat{\Yb}$ and infer, equivalently, a zero-prior-mean process
  $\Tempbhz-\mupr \sim \Gauss( 0 , \Gpr )$.  }
on $\Tempbhz$ by modeling $\Tempbhz$ 
as a discretized
solution of a stochastic PDE of the form
\begin{equation}  \label{eq:pdeGMRF}
\gamma \left(  \kappa^2 \mathcal{I} -\triangle   \right) \Temp(\sbo) = \mathcal{W}(\sbo), \qquad \sbo \in\Dom,
\end{equation}
where $ \mathcal{W}$ is a white noise process, $\kappa$ is a positive scalar parameter,
$\triangle$ is the Laplacian operator and $\mathcal{I}$ is the identity operator.  In
particular, we exploit the explicit link between Gaussian Markov random fields with the
Mat\'{e}rn covariance function and solutions to stochastic PDEs as outlined in
\cite{lindgren2011explicit}.  In this case, the action of a square root of the prior
covariance matrix on a vector is readily available as the solution of an elliptic PDE on
$\Dom$, and thus it is scalable to very large inverse problems
\cite{lindgren2011explicit}. 
In this example we use $h_c=23.8 {\rm \,W / m^2\,K}$ for the convective heat transfer
coefficient between the aluminum fin and the external fluid (air), which has constant temperature
$\Temp_{\infty} = 283\,{\rm K}$.  The width of the domain $\Dom$ in Figure
\ref{fig:heat_configuartion} is $H = 2 \times 10^{-2}\,{\rm m}$.
The heat flux $q(\sbo,t)$ is time-independent and nonnegative and can be written as the
superposition of two square impulse functions with zero background: one centered at
$6 \times 10^{-3} {\rm \,m}$ with width $8 \times 10^{-3} {\rm \,m}$ and intensity
$0.6 {\rm \, W / m^2}$; and the other centered at $15 \times 10^{-3} {\rm \,m}$ with width
$4 \times 10^{-3} {\rm \,m}$ and intensity $0.3 {\rm \, W / m^2}$.
Observations are collected every $\Delta t = 5 \times 10^{-4} {\rm \, s}$ for a total of
$M=100$ measurements.
We use $\sigma_{\text{obs}}= 1/2$ as the standard deviation of the
observational noise.
The prior parameters in \eqref{eq:pdeGMRF} are given by $\gamma = 1 \times 10^{4} $ and
$\kappa = \sqrt{8}/\rho_{\rm pr}$ with $\rho_{\rm pr}=H/10$.  This choice of $\kappa$
defines a prior with correlation values near $1/10$ at distance $\rho_{\rm pr}$
\cite{lindgren2011explicit}. The original prior mean is set to
$\mu_{\rm pr}=318\,{\rm K}$.  However, we equivalently infer the zero-prior-mean process
$\Tempbhz-\mupr$ as explained in the previous footnote. %

\subsection{Goal-oriented linear inverse problem}
\begin{figure}[!htp]
\begin{center}
\includegraphics[width=0.30\textwidth]{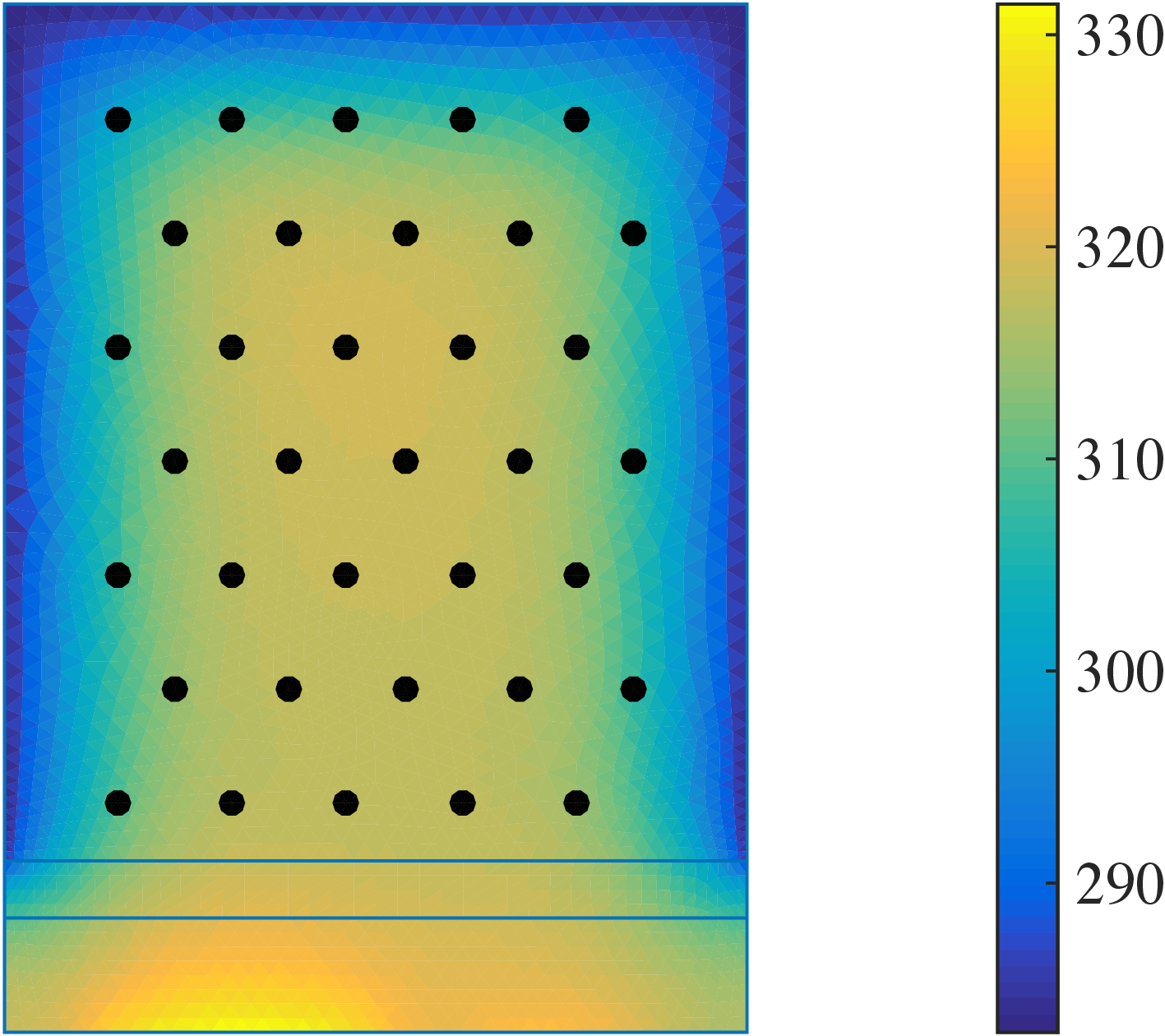}
\hspace{50pt}
\includegraphics[width=0.30\textwidth]{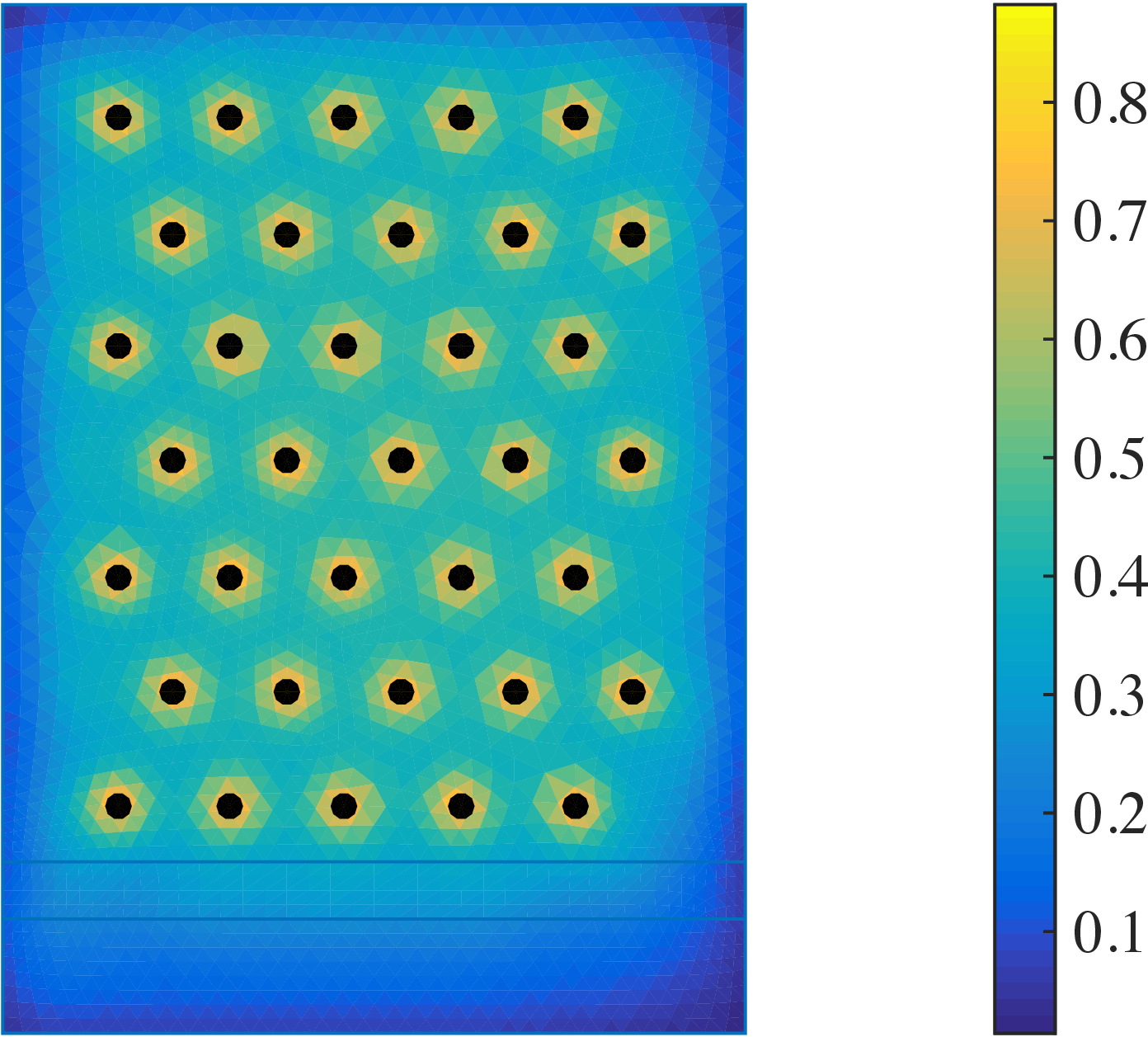}
\caption{\emph{(left)} Initial temperature field used to generate synthetic data according
  to the problem configuration described in Section \ref{s:examples}. This temperature
  field was not drawn from the prior distribution of $\Tempbhz$; instead, it corresponds
  to a finer discretization of the continuous stochastic process $\Temp$ evaluated at the
  initial time.  \emph{(right)} Normalized difference between the prior and posterior
  variance of the parameters, i.e.,
  $( \Var(\Tempbhz)-\Var(\Tempbhz \vert \Yb) ) / \,\Var(\Tempbhz)$.  The regions of
  greatest relative decrease of the variance are localized around the sensor locations
  (black dots).  }
\label{fig:heatGoalData} 
\end{center}
\end{figure}

We now introduce the goal-oriented feature of the problem.  As stated earlier, we are only
interested in the initial temperature distribution over the CPU (i.e., in $\Dom_1$).  Let
$\Zb$ be the restriction of $\Tempbhz$ to the domain of interest $\Dom_1$.  Clearly, there
is a linear map between $\Zb$ and $\Tempbhz$, i.e., $\Zb = \gop \, \Tempbhz$ with
$\gop \in \re^{p\times n}$ and $p \ll n$.  Thus, we have a linear-Gaussian goal-oriented
inverse problem as introduced in Section \ref{s:theory}:
\begin{equation}
  \begin{cases}
    \Yb = G \, \Tempbhz+\error\\
    \Zb = \gop \,\Tempbhz, 
  \end{cases}	
\end{equation}
where both the marginal distribution of $\Tempbhz$ and the likelihood $\Yb\vert\Tempbhz$
are specified. (In this example we denote the parameters by $\Tempbhz$ rather than $\Xb$.)  We choose
a finite element discretization of the temperature field such that $\Tempbhz\in\re^{2400}$
and $\Zb\in\re^{370}$.  Our goal is to characterize optimal approximations of the
posterior statistics of the QoI, $\Zb \vert \Yb$, for a given set of observations (see
Figure \ref{fig:heatGoalData} ({\it left})).  In this case, computing the posterior
distribution of the QoI using direct formulas like \eqref{eq:QoIstatistics} is infeasible
as the QoI is a finite-dimensional approximation to a distributed stochastic process,
$\Temp(0)|_{\Dom_{1}}$, and can be arbitrarily high-dimensional depending on the chosen
level of discretization.
The configuration of this problem highlights a crucial aspect of dimensionality reduction
of goal-oriented inverse problems.  Ideally we would place the sensors on $\Dom_{1}$ since
we are interested in inferring the temperature field on the CPU.  However, due to
geometrical constraints, we are forced to place our sensors on the heat sink ($\Dom_{3}$).
As a result, observations are much more informative about the parameters in $\Dom_{3}$
than in $\Dom_{1}$.  We see a hint of this in Figure \ref{fig:heatGoalData} ({\it right}),
which shows the normalized difference between the prior and posterior variance of the
parameters, $( \Var(\Tempbhz)-\Var(\Tempbhz \vert \Yb) ) / \Var(\Tempbhz)$.  The prior
variance is  reduced the most around the sensor locations in $\Dom_3$, which makes intuitive
sense as the data are increasingly less informative as we move away from the sensors.

\begin{figure}[!htp]
\begin{center}
\includegraphics[width=0.21\textwidth]{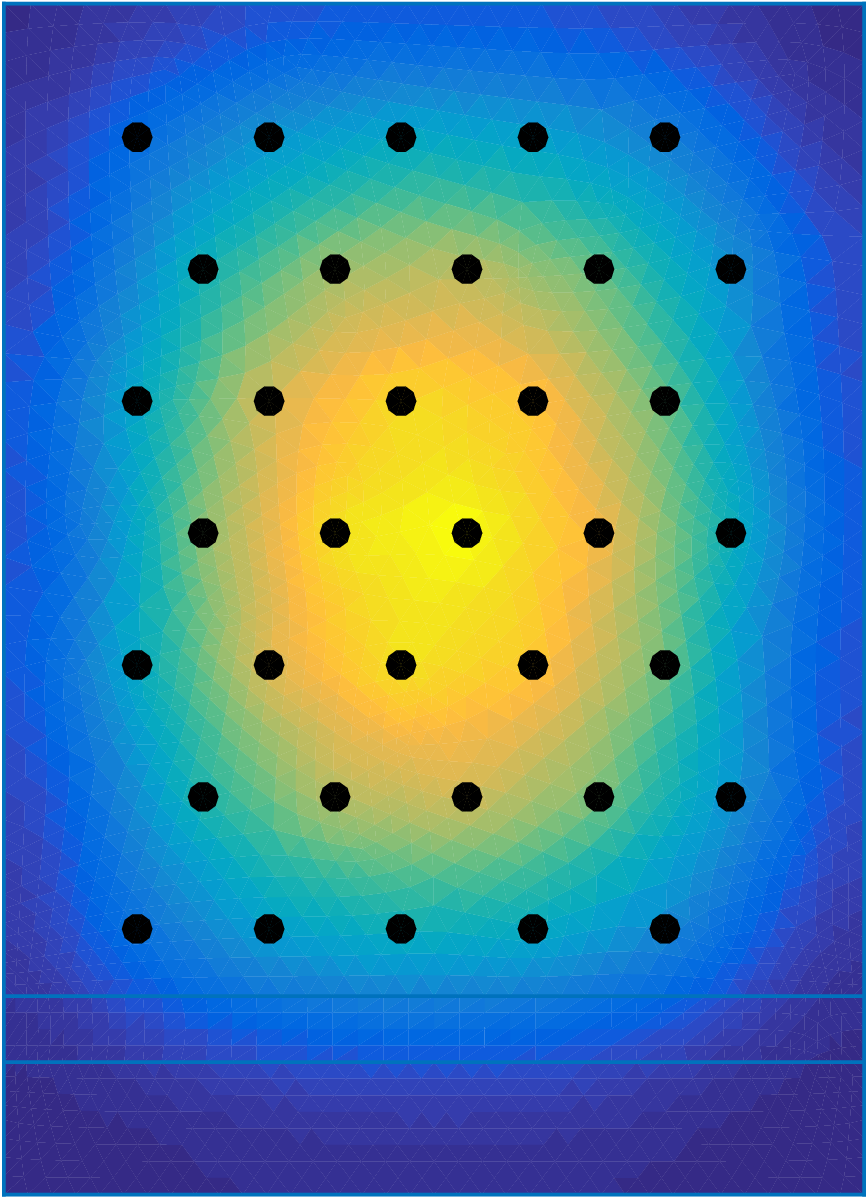}
\hspace{20pt}
\includegraphics[width=0.21\textwidth]{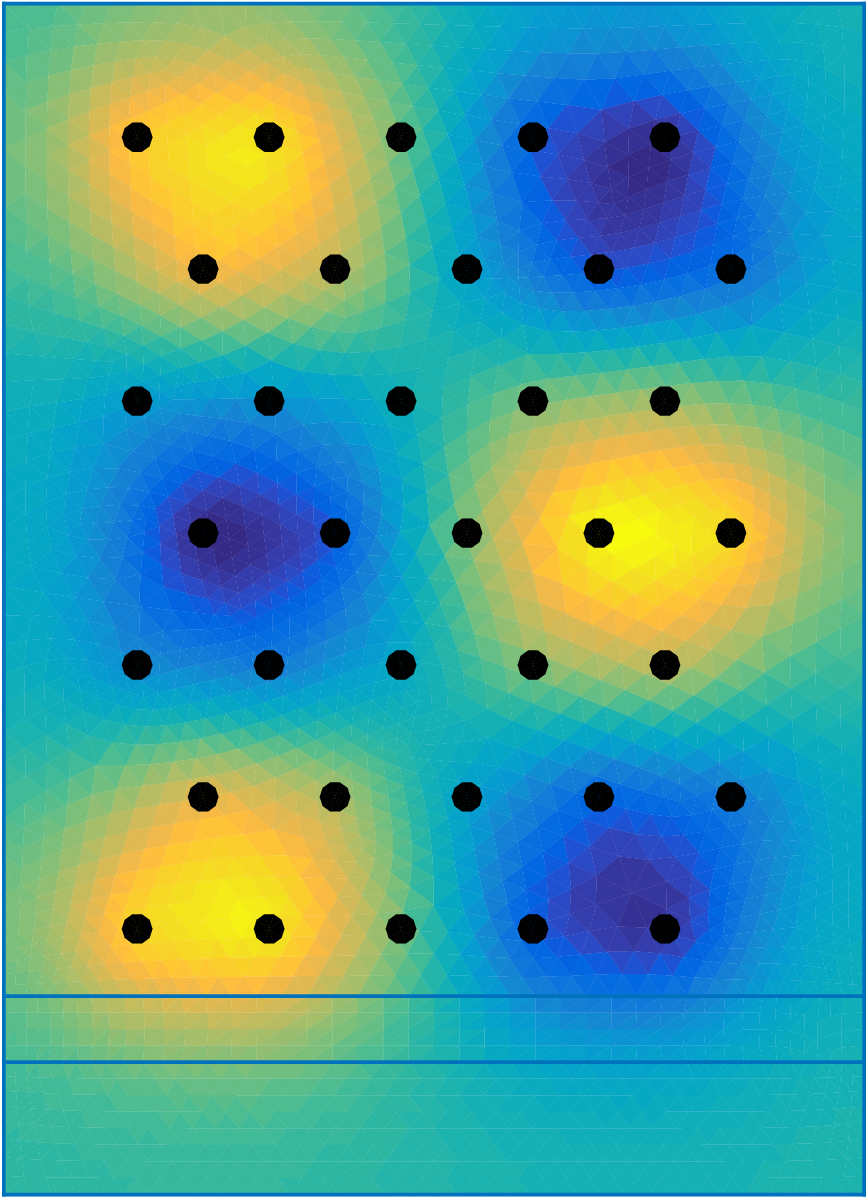}
\hspace{20pt}
\includegraphics[width=0.21\textwidth]{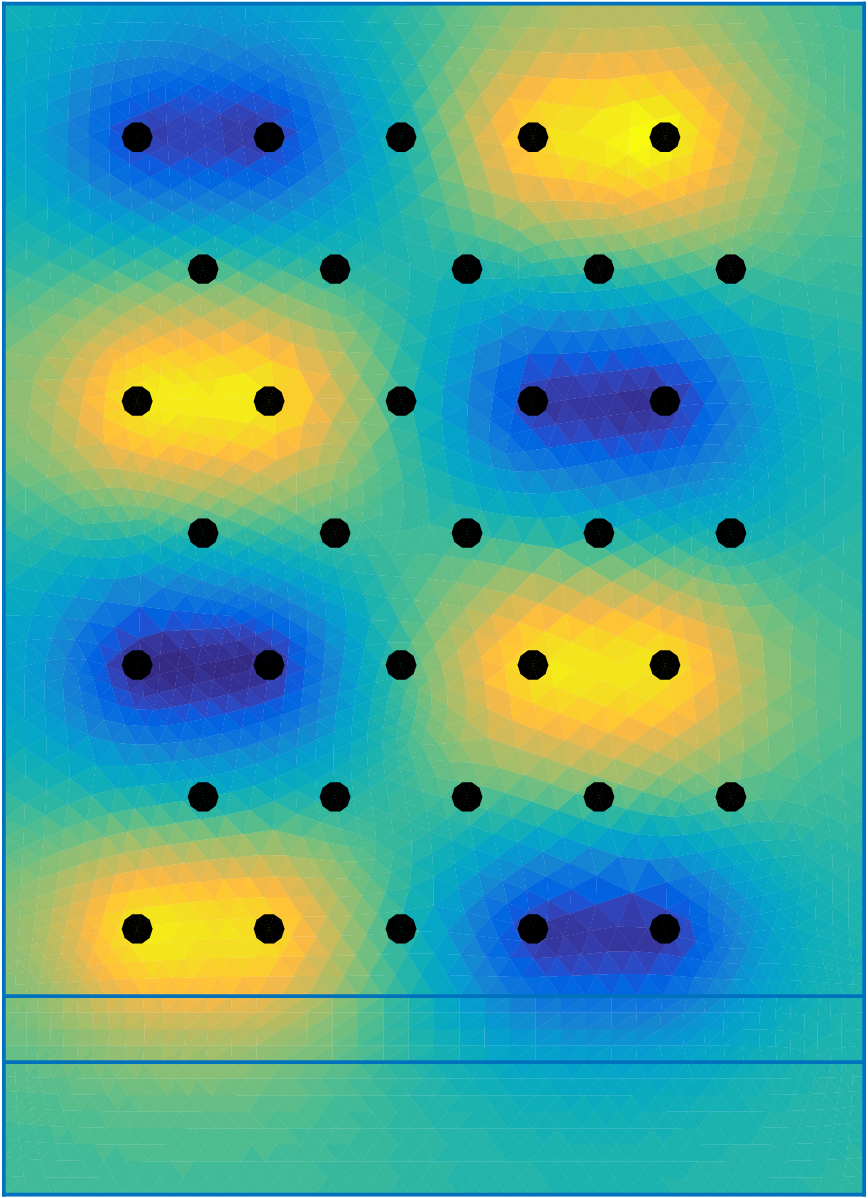}
\caption{Three eigenvectors $(w_i)$ of the matrix pencil
  $(H,\Gpr^{-1})$ as defined in Theorem \ref{thm:mainLinear}: $w_1$
  \emph{(left)}, $w_6$ \emph{(center}), and $w_{10}$ \emph{(right)}.
  These eigenvectors define the prior-to-posterior update in the
  optimal approximation \eqref{eq:optCovLinear} of the posterior
  covariance of the parameters $\Gpos$. Note that these leading
  eigenvectors have features near the locations of the sensors in
  $\Dom_3$.  This is the region where the data are most informative
  for the parameters, but not necessarily for the QoI.  }
\label{fig:eigvecs_pencil_linear} 
\end{center}
\end{figure}

\begin{figure}[!htp]
\begin{center}
\includegraphics[width=0.44\textwidth]{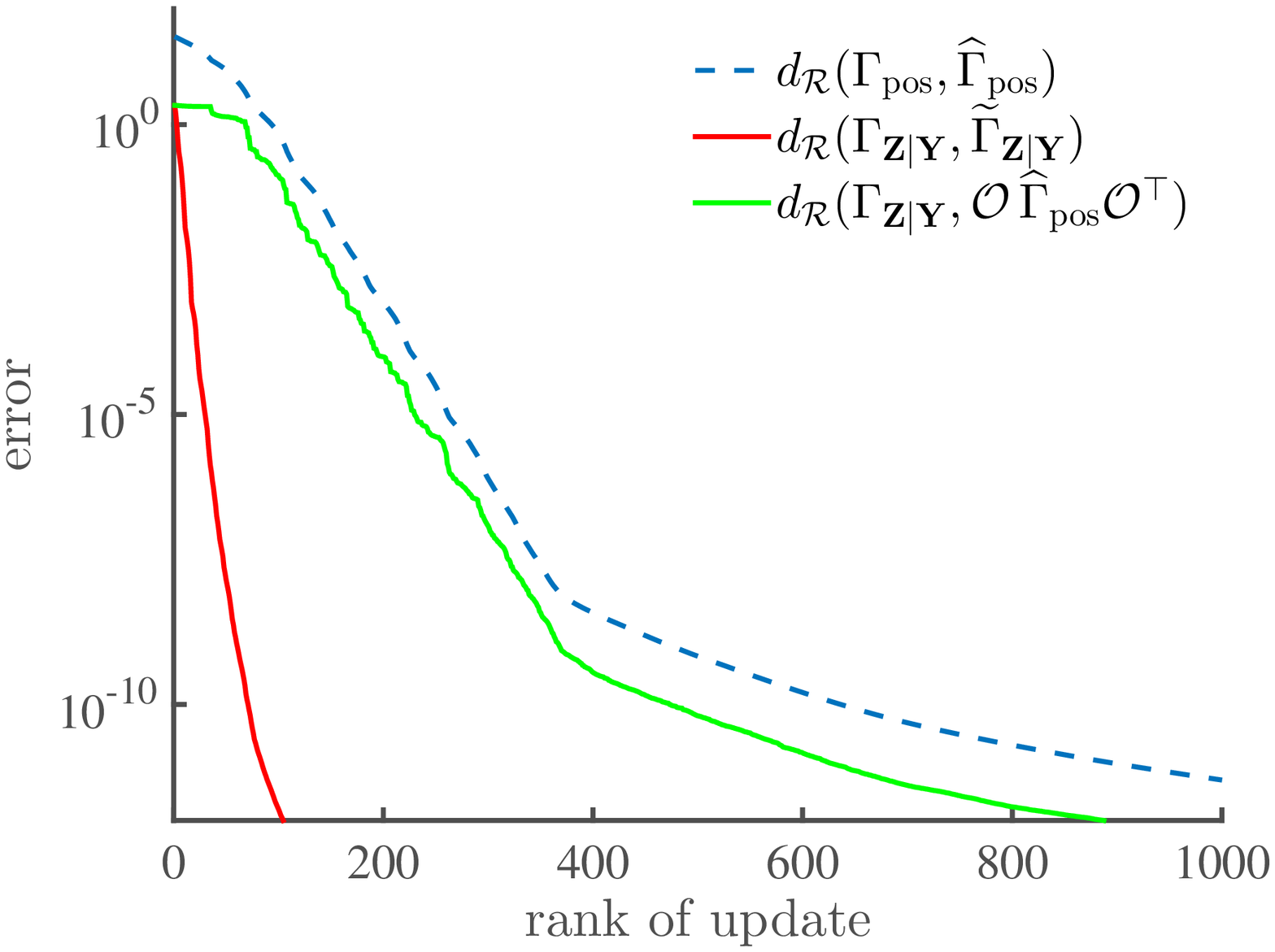}
\includegraphics[width=0.44\textwidth]{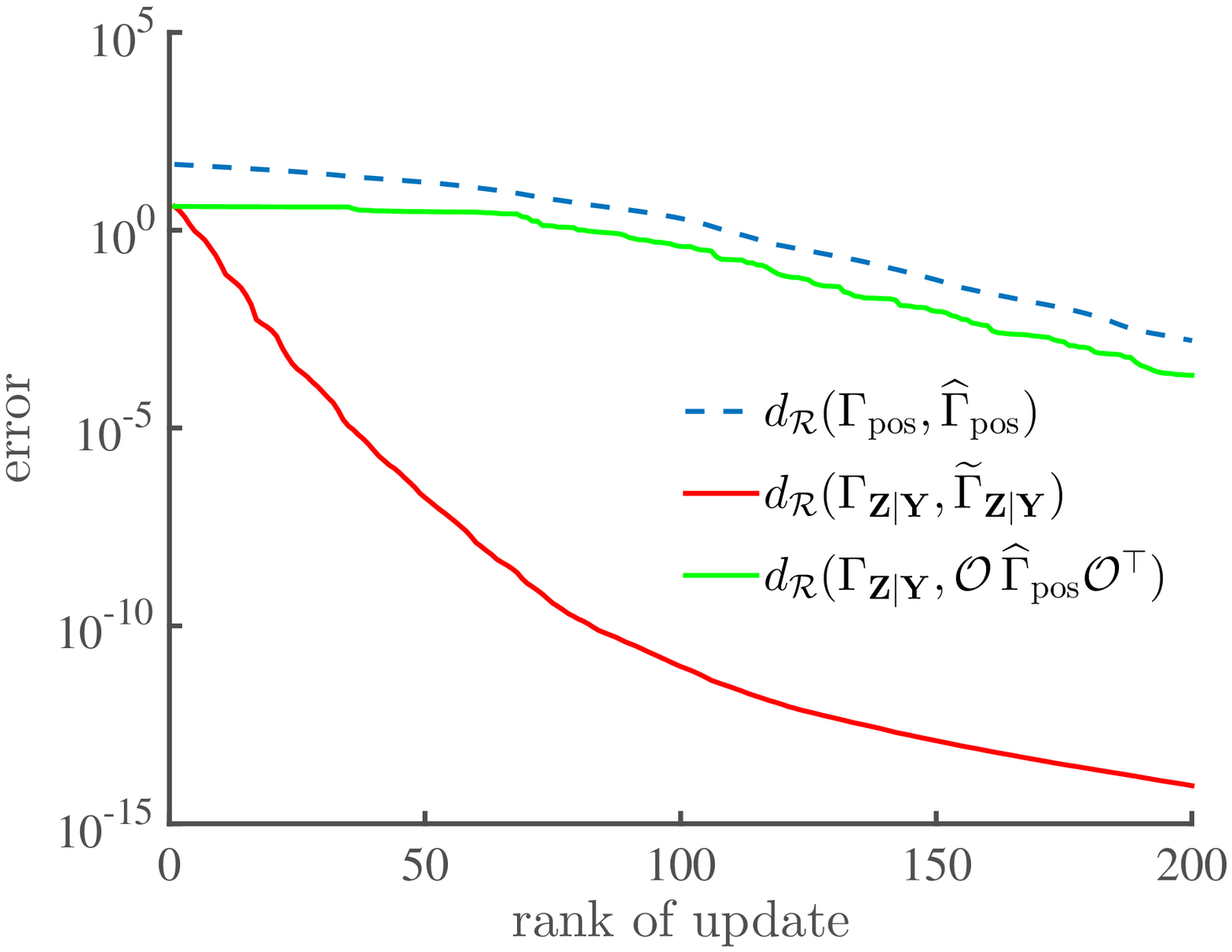}
\caption{\emph{(left)} Convergence of the covariance approximations in the natural
  geodesic distance over the manifold of SPD matrices (see Section \ref{s:metrics}). The
  blue dotted line shows the  distance between the covariance of the parameters
  $\Tempbhz|\Yb$ (i.e., $\Gpos$) and its optimal approximation $\Gposh = \Gpr - KK^\top$,
  as a function of the rank of  $K$ (see Theorem \ref{thm:mainLinear}). The red line shows
  the distance between $\Gzy$ and its optimal approximation introduced in Theorem
  \ref{approxCovUpdate-qoi},  $\Gzyopt = \Gz - KK^\top$, as a function of the rank of
  $K$. Finally, the green line shows the  distance between $\Gzy$ and the suboptimal
  approximation \eqref{eq:approxCovQoIsub} obtained as $\gop \, \Gposh \, \gop^\top$.
\emph{(right)} Detail of the figure on the left, with both axes rescaled.
}
\label{fig:eig1} 
\end{center}
\end{figure}

We first focus on the approximation of the posterior covariance of the QoI.  If we use the
suboptimal approximation introduced in \eqref{eq:approxCovQoIsub}, then we have to pay a
considerable computational price as a result of the data being informative about
directions in the parameter space that are not relevant to the QoI. This issue is
illustrated by the numerical results in Figure \ref{fig:eig1}.  To set the stage, we begin
with the posterior covariance $\Gpos$ of the parameters $\Tempbhz$ and construct the
optimal approximation $\Gposh=\Gpr - KK^\top$ from Theorem \ref{thm:mainLinear}. Though
this approximation is optimal for any given rank of the update, its convergence in this
problem is rather slow---as shown by the dotted blue line in Figure \ref{fig:eig1}---because
there are many data-informed directions in the parameter space. (Notice the multitude of
sensors on the heat sink in Figure \ref{fig:heat_configuartion}, each yielding
observations at $M$ successive times.)
If  we now use $\Gposh$ to yield an approximation of the actual posterior covariance of
interest $\Gzy$ by means of $\Gzy \approx \Gzysub = \gop \, \Gposh \, \gop^\top$ (i.e., the na\"{i}ve
approximation of \eqref{eq:approxCovQoIsub}), then the convergence of this approximation is
still slow, as seen in green solid line of Figure \ref{fig:eig1}.  This slow
convergence can be easily explained.  The optimal approximation $\Gposh$ of $\Gpos$
accounts first for those directions that are most informed by the data.  These directions
correspond to modes with features near the sensors in $\Dom_3$ (see Figure
\ref{fig:eigvecs_pencil_linear}), but they provide little information about the
parameters in the region of interest ($\Dom_1$).

\begin{figure}[!htp]
\begin{center}
\includegraphics[width=0.21\textwidth]{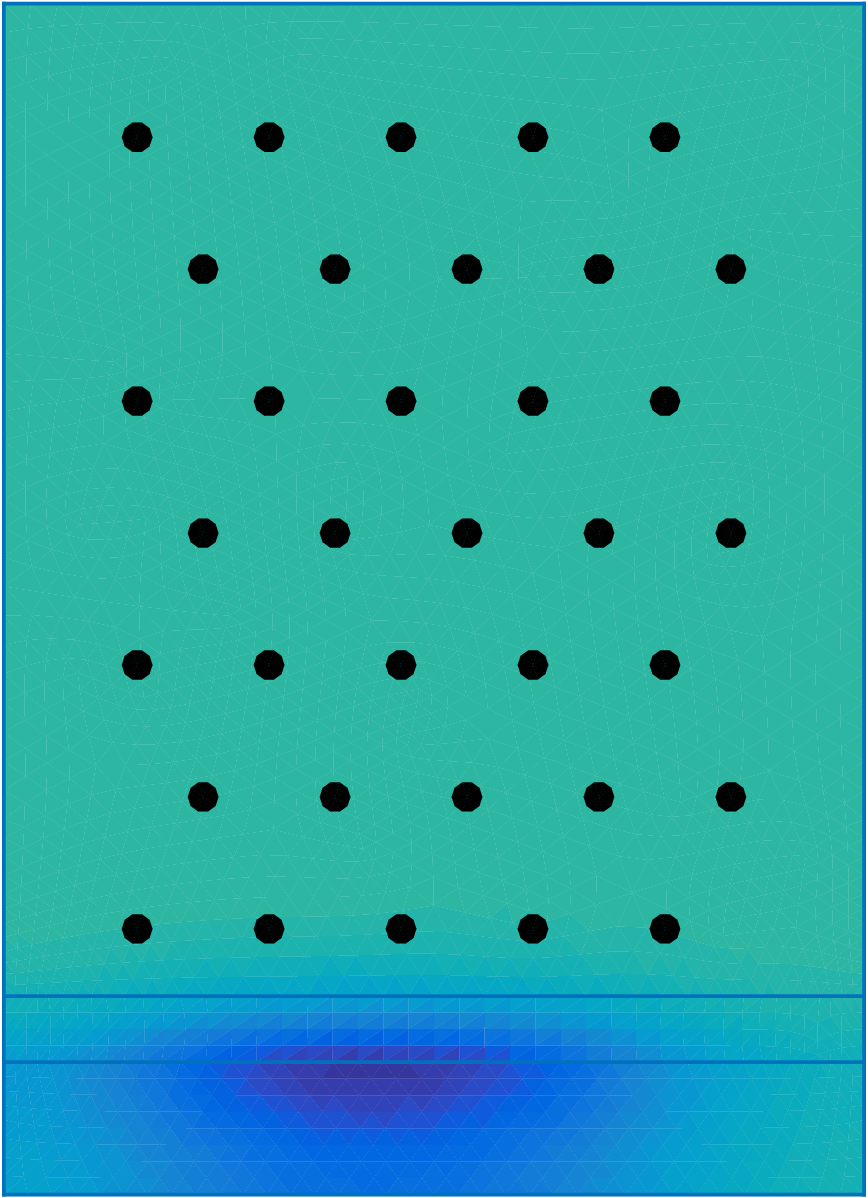}
\hspace{20pt}
\includegraphics[width=0.21\textwidth]{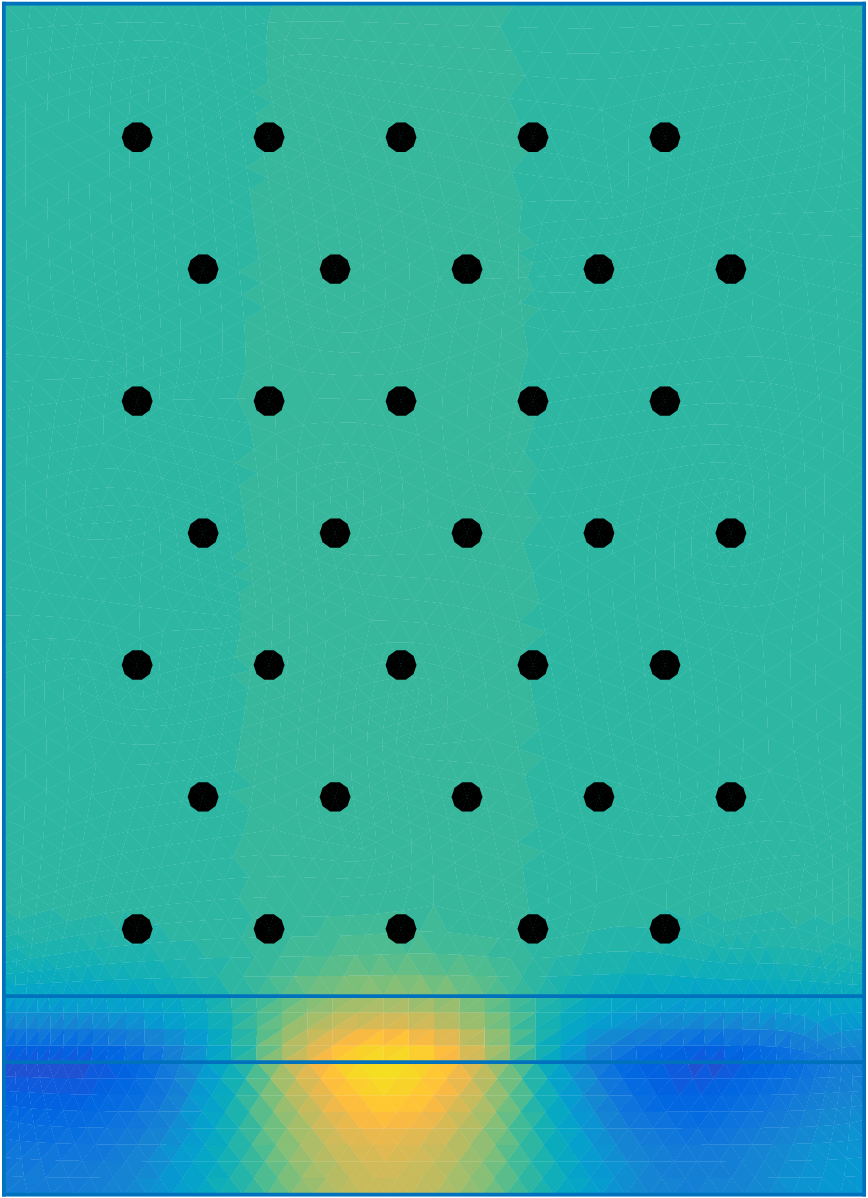}
\hspace{20pt}
\includegraphics[width=0.21\textwidth]{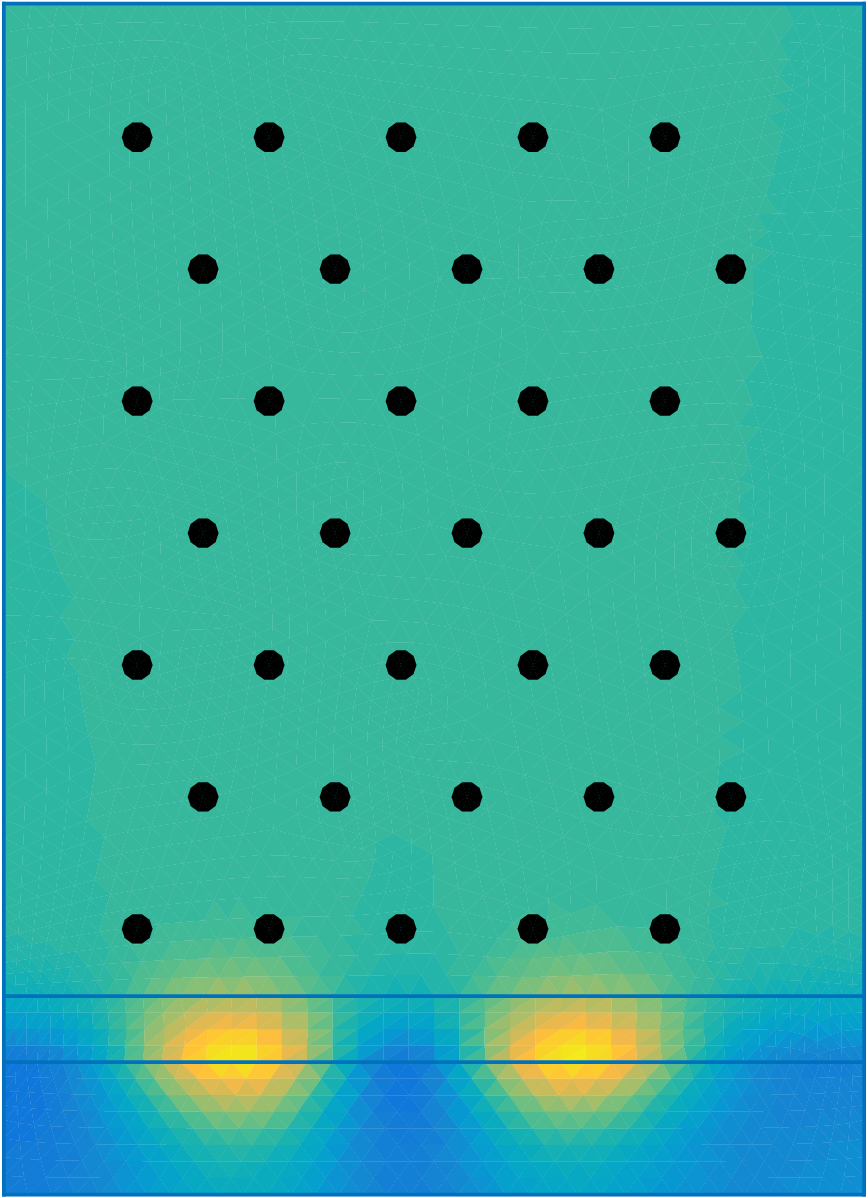}
\caption{Three vectors $(\widetilde{q}_i)$ defining the prior-to-posterior update in the
optimal \emph{goal-oriented} approximation of $\Gpos$ introduced in 
\eqref{eq:goalOrientedGpos} (see Lemma \ref{cor:optGoalApproxCov}).
In particular, we show $\widetilde{q}_1$ \emph{(left)},  
$\widetilde{q}_3$ \emph{(center)}, and 
$\widetilde{q}_5$ \emph{(right)}.
One can interpret these vectors as directions in the parameter space that are informed
by the data, relative to the prior, \emph{and} that are relevant to the QoI.
The relevant features of the  $(\widetilde{q}_i)$ are concentrated
around the region of interest $(\Dom_1)$.
These directions should be contrasted with the modes in Figure 
\ref{fig:eigvecs_pencil_linear},
which are strongly informed by the data but, at the same time, nearly irrelevant to the QoI.
}
\label{fig:eigvecs_pencil_goal} 
\end{center}
\end{figure}

On the other hand, if we use the optimal approximation of $\Gzy$ defined in Theorem
\ref{approxCovUpdate-qoi}, then convergence is remarkably fast, as illustrated via the
red solid line in Figure \ref{fig:eig1}. Now we only need to update $\Gz$ along a handful
of directions---say twenty---to achieve a satisfactory approximation of $\Gzy$.  The key
to achieving such fast convergence is to confine the inference to directions in the
parameter space that are most informed by the data, relative to the prior, \textit{and}
that are relevant to the QoI.
Moreover, these fundamental directions can be explicitly extracted from a \textit{goal-oriented}
approximation of the posterior covariance of the parameters, as explained in Lemma
\ref{cor:optGoalApproxCov}; three such directions are shown in Figure
\ref{fig:eigvecs_pencil_goal}.

We note that $\Gzy$ is by no means a low-rank matrix. (See its spectrum in
Figure \ref{fig:eigCovQoI} (\textit{left})).  This situation is fairly typical when
dealing with large-scale inverse problems with non-smoothing priors (e.g., Gaussian fields
with correlation function of Mat\'{e}rn type) and limited observations.  In these
situations, seeking an approximation of $\Gzy$ as a low-rank matrix would be
inappropriate; that is, classic dimensionality reduction techniques, e.g.,
Karhunen--Lo\`{e}ve reduction \cite{marzouk2009dimensionality,li2006efficient}, are
quite inefficient. Instead, low-dimensional structure lies in the change from prior to
posterior, due to the data being informative, relative to the prior, only about a
low-dimensional subspace of $\mathbb{R}^p$. This fact justifies the choice of the
approximation class $\Mc_r^{\Zb}$ for $\Gzy$ in \eqref{eq:classQoI}.  The efficiency of
the approximation class $\Mc_r^{\Zb}$ is also evident from the sharp decay of the red
curve in Figure \ref{fig:eig1}: only a handful of directions in the prior-to-posterior
\textit{update} are needed for a good approximation of $\Gzy$.

\begin{figure}[!htp]
\begin{center}
\includegraphics[width=0.7\textwidth]{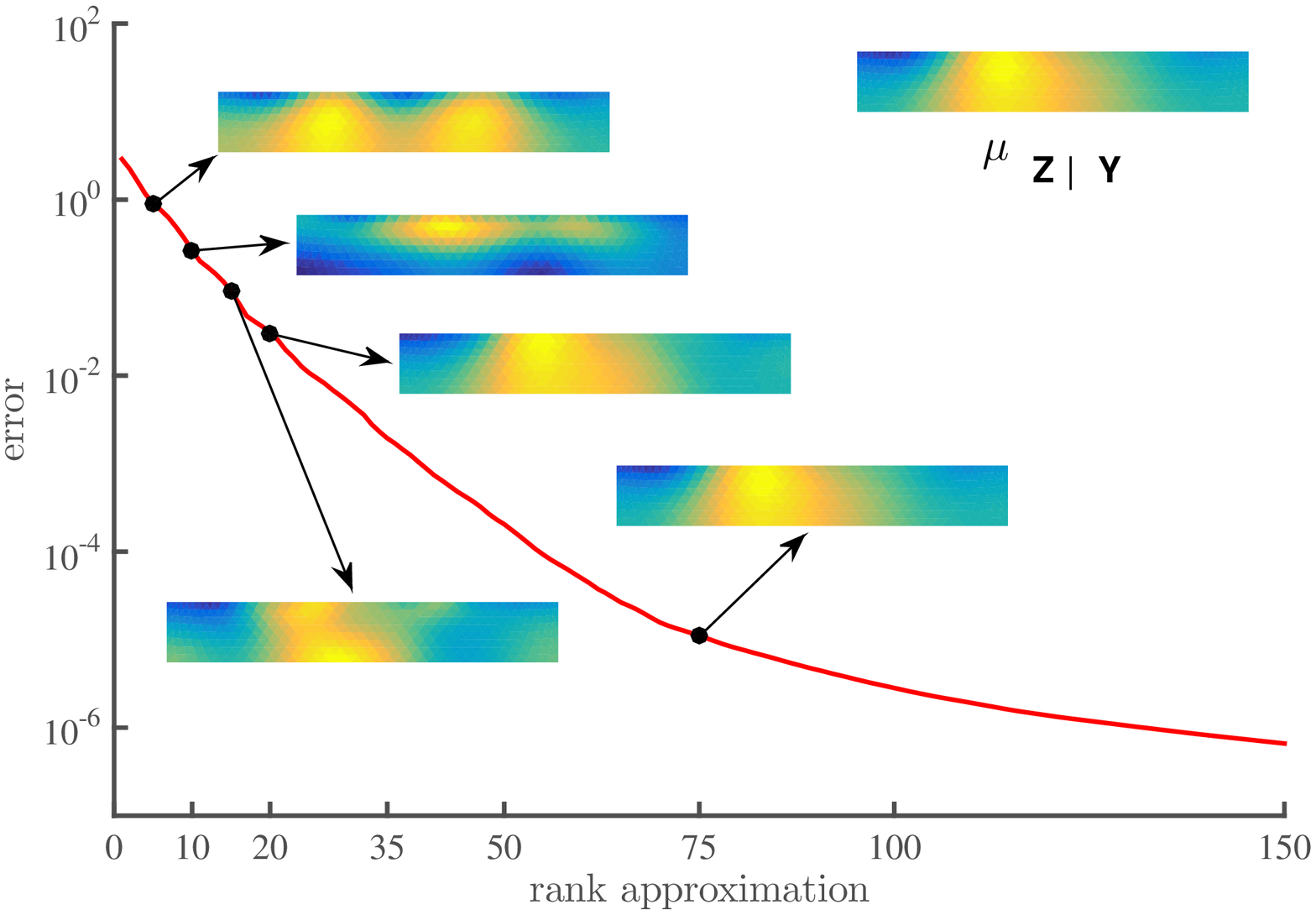}
\caption{The solid curve shows the error associated with the optimal low-rank
  approximation of the posterior mean of the QoI, $\muzy$, given in Theorem
  \ref{thm:mean_approx_lowrank}). The error is measured as the square root of $\Eb \left [ \left \Vert \muzy - A^* \Yb \right \Vert^2_{\Gzy^{-1}} \right ]$ and is a
  function of $\rank(A^*)$.
  The top right corner shows $\muzy$ for a particular realization of $\Yb$ (see Figure
  \ref{fig:heatGoalData} \emph{(left)}).  The snapshots along the solid curve show the
  corresponding approximation $\muzy\approx A^*\,\Yb$ for various ranks of $A^*$ and for
  the same realization of $\Yb$.  Notice that the approximation of the posterior mean of
  the QoI is already good with $\rank(A^*)=20$.}
\label{fig:errorMean} 
\end{center}
\end{figure}

The optimal approximation of the posterior mean of the QoI as a low-rank linear function
of the data, as introduced in Theorem \ref{thm:mean_approx_lowrank}, also converges very
quickly as a function of the rank of the approximation, as shown in Figure
\ref{fig:errorMean}.  Once a low-rank approximation of the form
\eqref{minimizerMeanLowRank} is available, then one can compute an accurate approximation
of $\muzy$ for each new realization of the data $\Yb$ by simply performing a low-rank
($r=20$ in this case) matrix-vector product.
See \cite{chung2014efficient,friedland2007generalized,markovsky2008structured,hua1998generalized,chung2012optimal,chung2011designing}
for a series of related efforts in a \textit{non}-goal-oriented but possibly non-Gaussian
framework.

\subsection{A nonlinear QoI} \label{sec:nonlinearQoI} 
\begin{figure}[!htp]
  \begin{minipage}[b]{0.50\linewidth}
    \centering
    \includegraphics[width=0.7\linewidth]{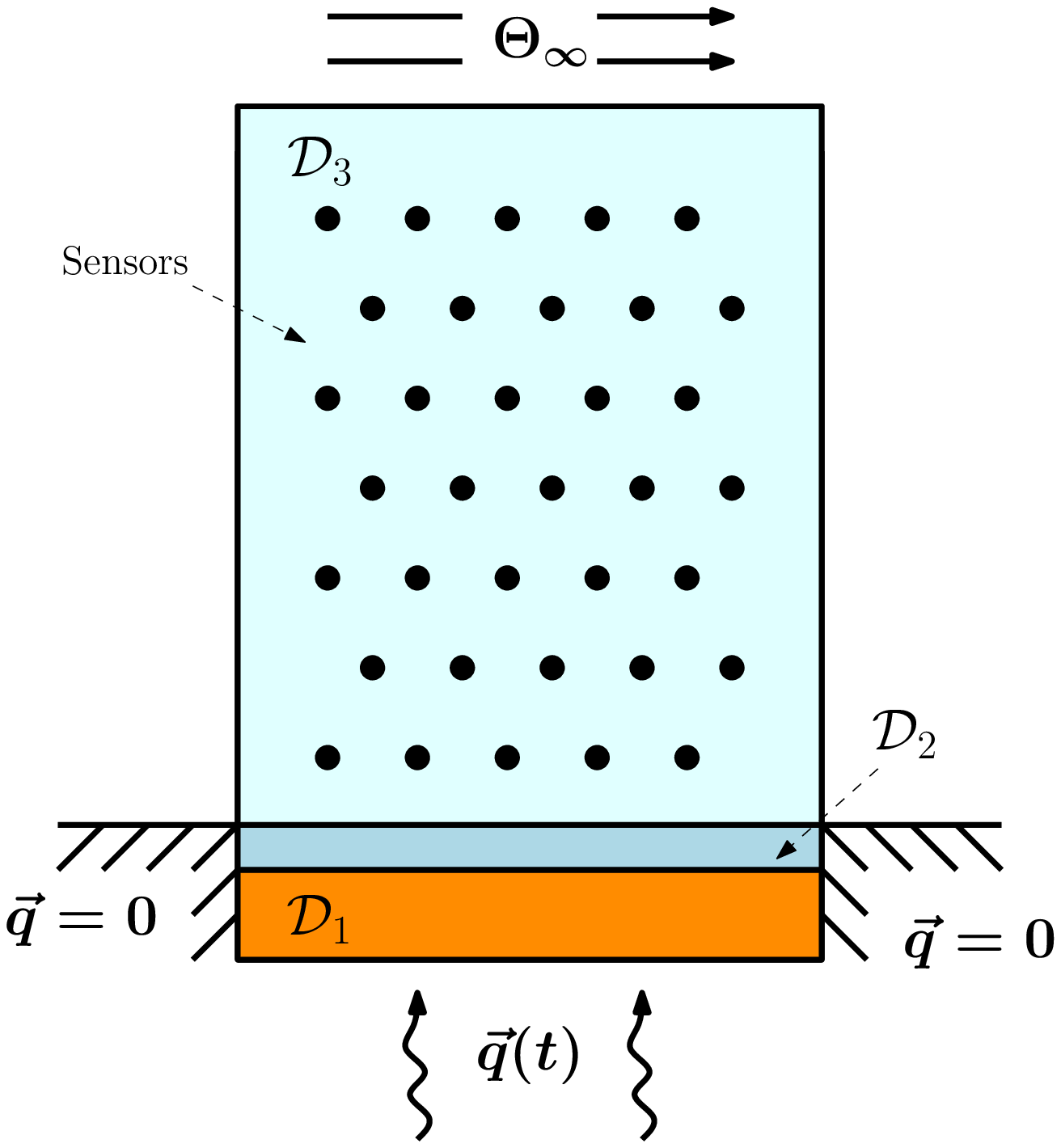}
    \par\vspace{0pt}
  \end{minipage}%
  \begin{minipage}[b]{0.50\linewidth}
    \centering
 	\begin{tabular}{
   c
   S[table-format = 1.2e-1]
   S[table-format = 1.2e1]
   S[table-format = 1.4]
 }
  \toprule
    \mc{Material} 
  & {$\alpha_i$ at \SI{20}{\celsius}}
  & {Domain} \\
  \midrule
    \mc{---}
  & {\si[per-mode = symbol]{\m\per\s^2}}
  & {---} \\
  \midrule
    Copper & 1.11e-4 & $\Dom_{1}$ \\
    Silicon &  8.8e-5 & $\Dom_{2}$ \\
    Aluminum   & 8.42e-5 & $\Dom_{3}$ \\
 			 \bottomrule
	\end{tabular}
	\par\vspace{65pt}
\end{minipage}

\caption{\emph{(left)} CPU cooling problem. Inversion for the initial temperature field on
  $\Dom_1$ given noisy temperature measurements on an aluminum heat sink ($\Dom_3$). The
  figure shows the problem configuration, the locations of the sensors (black dots), and
  the boundary conditions for the heat equation describing time evolution of the
  temperature field on the domain $\Dom \coloneqq \Dom_1 \cup \Dom_2 \cup \Dom_3$.
\emph{(right)} Material properties of the different layers.
}
\label{fig:heat_configuartion} 
\end{figure}

\begin{figure}[!htp]
\begin{center}
\includegraphics[width=0.44\textwidth]{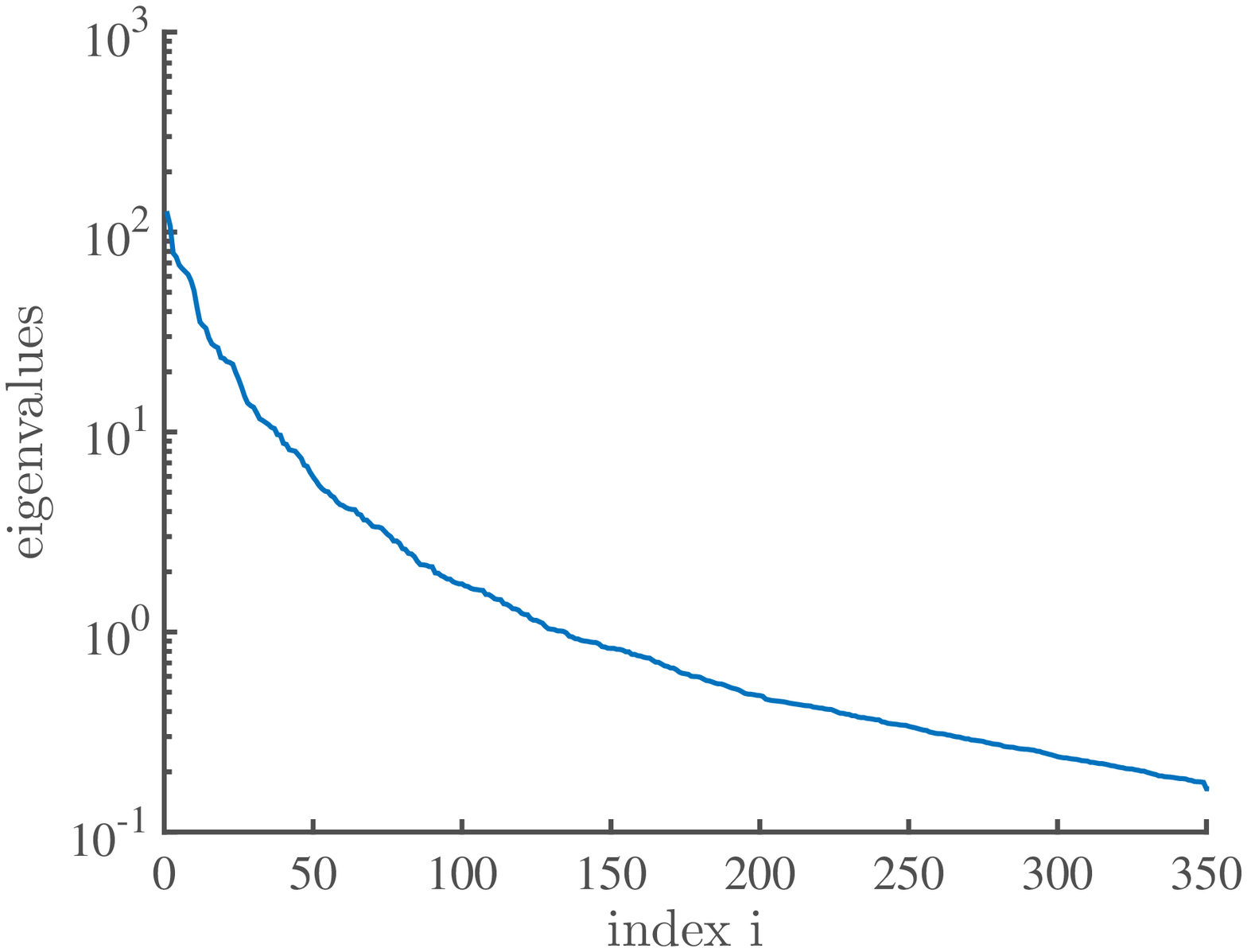}
\includegraphics[width=0.44\textwidth]{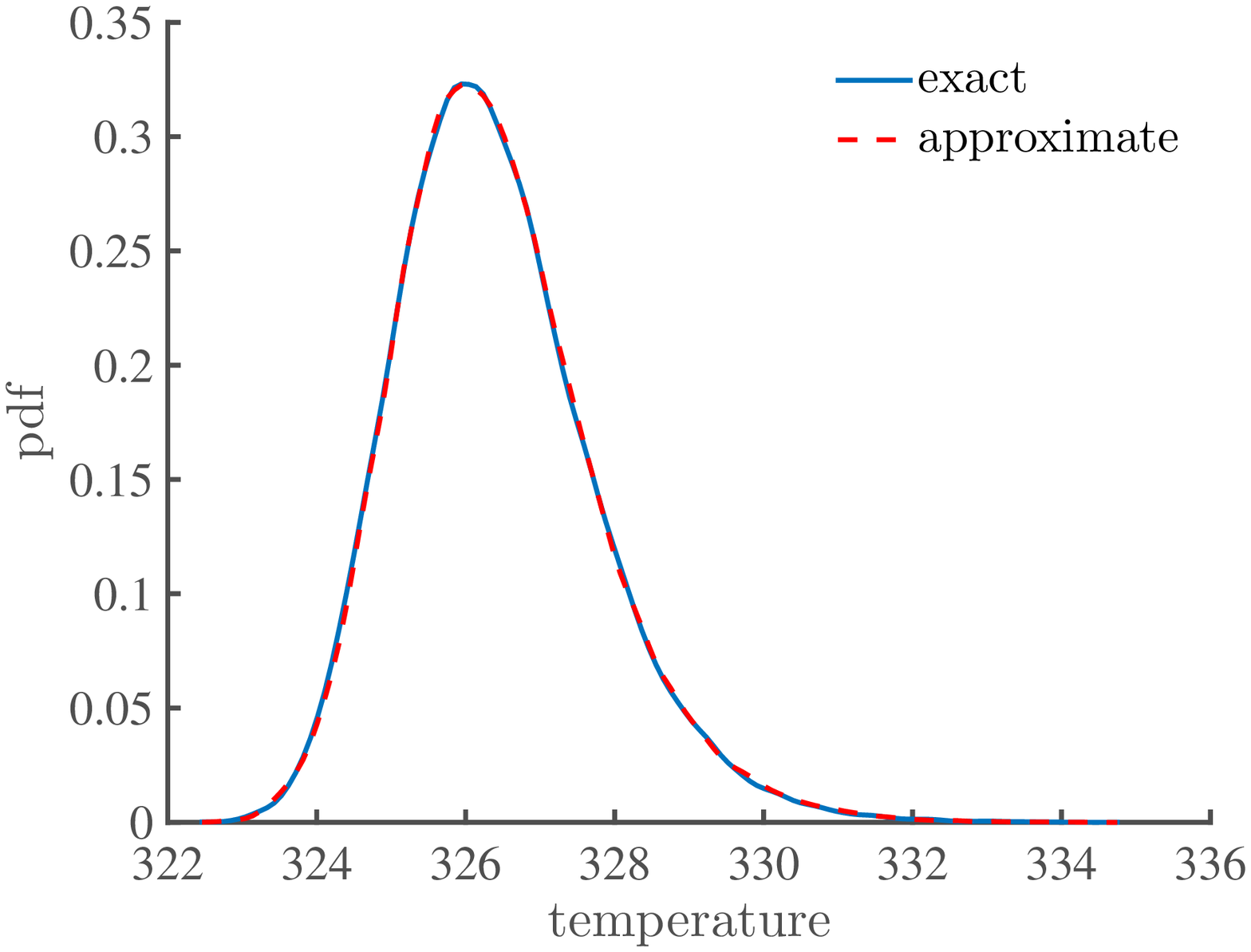}
\caption{ \emph{(left)} Eigenvalues of $\Gzy$.  For this problem configuration
  $\Zb\in\re^{370}$, so the matrix $\Gzy$ is not low-rank.  \emph{(right)} The blue solid
  curve shows a kernel density estimate (KDE) of the \emph{exact} posterior density of the
  nonlinear QoI $\Zbh\coloneqq \max(\Zb)$, i.e., the density of $\Zbh\vert\Yb$,
  constructed from $1\times 10^6$ samples. Notice that this density is non-Gaussian.  The
  red dotted curve shows a KDE constructed with $1\times 10^6$ samples from an \emph{approximation} of
  $\Zbh\vert\Yb$ obtained as follows: First we sample the approximate measure
  $\mzya \coloneqq \Gauss( \muzy , \Gzyopt )$ obtained from an optimal approximation,
  $\Gzyopt$, of $\Gzy$ as a $20$--dimensional low rank update of $\Gz$ (see Theorem
  \ref{approxCovUpdate-qoi}).  Then, we push forward these samples through the nonlinear
  goal-oriented operator $g:\re^p\ra\re$.  The quality of the density approximation is
  already good for a rank--$20$ update.  These results are consistent with the theoretical
  bounds \eqref{eq:boundHell}. (See, in particular, the error curves in Figure
  \ref{fig:eig1}.)}
\label{fig:eigCovQoI} 
\end{center}
\end{figure}

We conclude this section by applying the approximation formulas introduced in this paper
to a particular case of nonlinear goal-oriented inference.
Suppose that we are only interested in the posterior distribution of the \textit{maximum}
temperature over $\Dom_1$ (see Figure \ref{fig:heat_configuartion}).  This is a useful QoI
because the material properties of a sensitive component (e.g., the CPU) might deteriorate
above a certain critical temperature (e.g., \cite{branco1996mechanical}).  In this case,
the QoI $\Zbh \coloneqq \max_{\Dom_1}\Tempbhz $ is a low-dimensional (in fact
scalar-valued) nonlinear function of the parameters. In general, let us write
$\Zbh=\nop(\Tempbhz)$ for some nonlinear function $\nop:\re^n \ra \re$.  Then we can cast
the nonlinear goal-oriented Bayesian inverse problem as
\begin{equation} \label{eq:protNonlinearQoI}
  \begin{cases}
    \Yb = G \, \Tempbhz+\error\\
    \Zbh  = \nop(\Tempbhz )
  \end{cases}	
\end{equation}
and try to characterize the posterior $\Zbh\vert \Yb$ for a particular realization of the
data. This problem is nontrivial, however, as $\Zbh\vert \Yb$ is non-Gaussian and cannot
easily be characterized by just two moments.
\hrevone{
In the most general case, one needs to resort to sampling techniques such as MCMC \cite{higdon2011posterior}
to characterize $\Zbh\vert \Yb$, 
or perhaps some deterministic alternative \cite{el2012bayesian,schillings2013sparse, dick2016higher,nagel2016spectral}.} 
Unfortunately, it is still not well understood how to properly adapt 
\hrevone{these} %
techniques to exploit ultimate goals and bypass full inference
of the parameters (see the offline--online strategy of \cite{lieberman2014nonlinear} for a
related effort in the context of goal-oriented nonlinear Bayesian inference).  Though
developing computationally efficient techniques to tackle general problems like
\eqref{eq:protNonlinearQoI} is of fundamental importance, in this particular example we
can adopt a much simpler, yet effective, approach.  Using the notation of this section,
notice that the nonlinear QoI $\Zbh$ can be written as $\Zbh=g(\Zb)$, where $\Zb$
represents the inversion parameters in the region of interest $\Dom_1$, and where
$g(\xb)=\max_i(x_i)$ for all $\xb=(x_1,\ldots,x_p)$.  Thus we can rewrite
\eqref{eq:protNonlinearQoI} as:
\begin{equation} \label{eq:protNonlinearQoI_equiv}
  \begin{cases}
    \Yb = G \, \Tempbhz+\error\\
    \Zb = \gop\,\Tempbhz \\
    \Zbh  = g(\Zb).
  \end{cases}	
\end{equation}
Then we can approximate the Gaussian posterior distribution of $\Zb\vert\Yb$ using the
goal-oriented techniques presented in this paper, and finally we can push forward the
latter distribution through the nonlinear operator $g$ to obtain a suitable approximation
of $\Zbh\vert\Yb$. 
\hrevone{ 
The nonlinear operator $g$ is never approximated in this process.}
That is, we first compute the posterior mean and the optimal
goal-oriented approximation of the covariance of $\Zb\vert\Yb$ using the results of
Theorem \ref{approxCovUpdate-qoi}, then we sample from the optimal approximating measure
$\mzya \coloneqq \Gauss( \muzy , \Gzyopt )$ using the results of Lemma \ref{cor:sqrtGyz},
and finally we push forward these samples through $g:\re^p\ra\re$ to obtain approximate
samples from the posterior distribution of the nonlinear QoI. 
 We can easily estimate the
{\it quality} of these approximate posterior samples using bounds like
\eqref{eq:boundHell}. 
\hrevone{ 
Note, however, that a bound like \eqref{eq:boundHell} quantifies only the
accuracy of the posterior moments, and does not yield an explicit measure
of distance between the non-Gaussian posterior distribution of $\Zbh\vert \Yb$ and its corresponding approximation.
Nevertheless, the plot on the right of Figure \ref{fig:eigCovQoI} shows that the
resulting approximation of the density of $\Zbh\vert\Yb$ is \hrevone{indeed quite good
for this particular choice of nonlinear operator $g$}.
}

\section{Conclusions}
\label{s:conclusions}
We have developed statistically optimal and computationally efficient approximations of
the posterior statistics of a \emph{quantity of interest} (QoI) in a goal--oriented
linear--Gaussian inverse problem. The posterior covariance of the QoI is approximated as a
low-rank negative update of the prior covariance of the QoI. Optimality holds with respect
to the natural geodesic distance on the manifold of symmetric positive definite matrices.
The posterior mean of the QoI is approximated as a low-rank function of the data, and
optimality follows from the minimization of the Bayes risk for squared-error loss weighed
by the posterior precision matrix of the QoI.  The minimization of this Bayes risk is
associated with the minimization of a Riemannian metric averaged over the distribution of
the data.
These optimal approximations avoid computation of the full posterior distribution of the
parameters and focus only on directions in the parameter space that are informed by the
data \emph{and} that are relevant to the QoI.  These directions are obtained as the leading
generalized eigenvectors of a suitable matrix pencil, and reflect a balance among all
the ingredients of the goal--oriented inverse problem: prior information, the forward model,
measurement noise, and the ultimate goals.  

An important avenue for future work is the extension of these
optimality results 
\hrevone{ 
to the case of nonlinear forward operators}. Here, we expect that
interpreting the QoI posterior approximation as the result of
composing the forward model with a carefully chosen projection
operator, as in \cite{cui2014likelihood}, may be quite helpful.
\hrevone{ 
Relaxing the Gaussianity assumptions on both the prior distribution and the measurement
noise are also important generalizations of the present work.}

\section*{Acknowledgments}
We would like to thank Olivier Zahm, Jayanth Jagalur,
Antoni Musolas, Ahmed Sameh, and Alicia Klinvex for many insightful
discussions and for pointing us to key references in the
literature. We would also like to thank Pierre-Antoine Absil, Omar Ghattas, and Georg Stadler
for many helpful remarks on computational issues. This work was
supported by the US Department of Energy, Office of Advanced
Scientific Computing (ASCR), under grant number DE-SC0009297.

\appendix
\section{Rao's metric between distributions}
\label{sec:diffGeom}
Let $\mathscr{M}=\{ \pi_\theta , \theta\in\Theta \}$ be a parametric family
of probability densities indexed by 
$\theta=(\theta_1,\ldots,\theta_n)\in\Theta$ \cite{atkinson1981rao}.
Rao considered a quadratic differential form given by
\begin{equation} \label{eq:quadFormRao}
	{\rm d}s^2 = \sum_{i,j} \,g_{ij}(\theta)\,{\rm d}\theta_i
        \,{\rm d}\theta_j \, ,
\end{equation}
where $g_{ij}(\theta)=\Ex_{\pi_\theta}[\,\partial_{\theta_i}\ln \pi_\theta \,
\partial_{\theta_j}\ln \pi_\theta\,]$ are the
entries of the Fisher information matrix, with $\Ex_{\pi_\theta}$ 
denoting integration with respect to $\pi_\theta$ \cite{fisher1922mathematical}.
The Fisher information matrix is a central object in mathematical statistics 
(e.g., the Cram{\'e}r-Rao inequality \cite{rao1945info}).
Intuitively, we can interpret \eqref{eq:quadFormRao} as the variance of the
function that describes the first order relative difference between $\pi_\theta$ and
a contiguous density, $\pi_{\theta + {\rm d}\theta}$, on $\mathscr{M}$ \cite{rao1949distance}.
The definition of a quadratic form like \eqref{eq:quadFormRao} 
allows us to measure curves on $\mathscr{M}$.
Given a smooth curve $\gamma:[0,1] \ra \Theta \simeq \mathscr{M} $, we can define its length as
$\ell(\gamma)\coloneqq \int_0^1 
(\sum_{i,j} \,g_{ij}(\gamma(t))\,{\rm d}\gamma_i  \,{\rm d}\gamma_j )^{1/2} 
\,{\rm d}t$ \cite{do1992riemannian}.
Thus, Rao's distance between a pair of distributions on $\mathscr{M}$ is simply
their geodesic distance, i.e., the length of the minimum length curve joining these distributions
\cite{rao1949distance}.
The quadratic form defined by the Fisher information matrix is invariant under
regular reparameterizations of $\mathscr{M}$ \cite{rao1945info}.
Thus, this fundamental invariance is also shared by Rao's distance which yields an 
intrinsic way of comparing distributions on $\mathscr{M}$.
Of course, it is possible to consider more general quadratic differential forms not based on the notion of Fisher information.
See \cite{rao1987differential} for various examples of differential metrics derived from
entropy functions or divergence measures
between probability distributions.
See \cite{amari2007methods} for a modern treatment of information geometry, the
field at the intersection of statistics and differential geometry.

\medskip
\section{Proofs of the main results}
\label{sec:proofs}
The following two Lemmas, \ref{lem:schur} and \ref{lem:linearModelQoi},
 will be used to prove  Theorems \ref{approxCovUpdate-qoi} and
\ref{thm:mean_approx_lowrank}.
We start with a result describing the relationships between different eigenpairs of the
Schur complements of a particular class of covariance matrices that arises in
Bayesian inverse problems.
\medskip

\begin{lemma}[Eigenpairs of Schur complements]\label{lem:schur}
Let $\Sigma\succ0$ be a matrix partitioned as
\begin{equation}
\Sigma = \begin{pmatrix}
 A & B\\
 B^\top & C
\end{pmatrix},
\end{equation}
where $A$ and $C$ are square matrices and $B\neq 0$. 
Then, $A$,$C$, and the Schur complements, $\Sc(A) \coloneqq C - B^\top A^{-1} B$ and
 $\Sc(C) \coloneqq A -   B C^{-1} B^\top$, are also SPD matrices.
Moreover:
\smallskip

\begin{enumerate}
\item \label{lemSchurPre}
If $(\beta , w )$ is an eigenpair of $( B C^{-1} B^\top , A )$, then 
$\beta < 1$ and $(1 - \beta ,  w)$ is an eigenpair of $(\Sc(C), A)$. Furthermore,
if $\beta \neq 0$, then
$((1-\beta)^{-1},B^\top w)$ is an eigenpair of $(\Sc(A)^{-1}, C^{-1})$.
 \smallskip

\item  \label{lemSchurMain} If $\beta \neq 0$ and $(\beta , w )$ is an eigenpair of  
$( B C^{-1} B^\top , A )$, then
$(\beta(1-\beta)^{-1},B^\top w)$ is an eigenpair of 
$(C^{-1} B^\top \Sc(C)^{-1} B C^{-1},C^{-1})$.
 \smallskip

\item \label{lemSchurLI}  
If $w_1,\ldots,w_k$ are linearly independent eigenvectors of  
$( B C^{-1} B^\top , A )$ with associated
eigenvalues $\beta_1\geq \beta_2\geq \cdots\geq \beta_k>0$, then $B^\top w_1,\ldots,B^\top w_k$ are linearly independent.
Moreover, if $k = {\rank}(B C^{-1} B^\top)$, then 
there can be at most $k$ linearly independent eigenvectors
of $(C^{-1} B^\top \Sc(C)^{-1} B C^{-1},C^{-1})$ associated with 
strictly 
positive eigenvalues.
\end{enumerate}
\end{lemma}
\medskip

\begin{proof}
The fact that 
 $A$, $C$, $\Sc(A)$ and
 $\Sc(C)$ are SPD matrices
 follows from  \cite{boyd2004convex}.
(\ref{lemSchurPre}) 
From $B C^{-1} B^\top w = \beta A w$ we obtain 
$\Sc(C)w = (1-\beta)A w$, which also implies that $\beta<1$ as
$\Sc(C)\succ 0$ and $A\succ 0$.
If $\beta \neq 0$, then $B^\top w \neq 0$ and
\begin{eqnarray*}
\Sc(A)^{-1} B^\top\,w & = & [\, C^{-1} + C^{-1} B^\top \Sc(C)^{-1} B C^{-1}\,]\,B^\top\, w\\
&=& [\,C^{-1} B^\top + C^{-1} B^\top \Sc(C)^{-1} (A-\Sc(C))\,] w\\
&=& (1-\beta)^{-1}\,C^{-1} B^\top \,w .
\end{eqnarray*}
where we used the Woodbury identity to rewrite $\Sc(A)^{-1}$.
(\ref{lemSchurMain}) It follows from (\ref{lemSchurPre}) that $( (1-\beta)^{-1} , B^\top w )$ is an
eigenpair of $( \Sc(A)^{-1} , C^{-1} )$ and by
$\Sc(A)^{-1}  = C^{-1} + C^{-1} B^\top \Sc(C)^{-1} B C^{-1}$
that
$(\beta (1-\beta)^{-1} , B^\top w )$ is an eigenpair 
of 
$(C^{-1} B^\top \Sc(C)^{-1} B C^{-1},C^{-1})$.
(\ref{lemSchurLI}) If $\sum_{j=1}^k a_j B^\top w_j =0$, then 
$A\sum_{j=1}^k \beta_j a_j w_j =0$, and therefore
$\sum_{j=1}^k \beta_j a_j w_j =0$ since $A\succ 0$, which leads to
$\beta_j a_j=0$ for $j=1,.\dots,k$ since $(w_j)$ are 
linearly independent, and thus $a_j=0$ for $j=1,.\dots,k$ since $\beta_j>0$.
Moreover, notice that
$\rank(C^{-1} B^\top \Sc(C)^{-1} B C^{-1})=\rank(B^\top \Sc(C)^{-1} B)=
{\rank}(B C^{-1} B^\top)$. 
Thus, there can be at most $\rank(B C^{-1} B^\top)$ linearly
independent  eigenvectors
of $(C^{-1} B^\top \Sc(C)^{-1} B C^{-1},C^{-1})$ with 
nonzero eigenvalues.
\end{proof}
\smallskip

{\bf Proof of Lemma \ref{lem:linearModelQoi}.} 
 Consider the identity
 $\Yb = G\,\Xb + \error = G \, \gopd \,\gop\, \Xb + 
 G\,( I - \gopd \, \gop) \,\Xb + \error =  G \,\gopd\, \Zb + \errorTr$,
where 	$ \gopd \coloneqq  \Gpr \gop^\top \Gz^{-1}$ and 
$\errorTr \coloneqq G\,( I - \gopd \,\gop) \,\Xb + \error $.
A simple computation shows that $\Ex[ ( I - \gopd \, \gop) \Xb\, \Zb^\top ] =0$.
Hence, $( I - \gopd \, \gop) \Xb$ and $\Zb$ are uncorrelated, and, more importantly, independent since they are also jointly Gaussian. It follows that $\errorTr$ and $\Zb$ are also independent since $\error$ was independent of $\Xb$ and $\Zb=\gop\,\Xb$. In the hypothesis of zero prior mean, the mean of $\errorTr$ is also zero. 
Moreover, $\Gamma_{\errorTr}  = \Var[  G\,( I - \gopd \,\gop) \Xb ]  + \Var[ \error]$
since $\Xb$ and $\error$ are independent. 
Simple algebra leads to the particular form of $\Gamma_{\errorTr}$.
\hfill $\square$
\smallskip

We can now prove the main results of this paper.
\smallskip

{\bf Proof of Theorem \ref{approxCovUpdate-qoi}.} 
By applying \cite[Theorem 2.3]{spantini2014optimal} to the linear Gaussian model defined in Lemma 
\ref{lem:linearModelQoi}, we know that   
a minimizer, $\Gzyh$, of the geodesic distance, $\dr$,  between $\Gzy$ and an element of 
${\Mc}_r^{{\Zb}}$ is given by:
$\Gzyh=\Gz-\sum_{i=1}^r \eta^2_i (1+\eta_i^2)^{-1} \, \widehat{q}_i \widehat{q}_i^\top$,
where $(\eta_i^2, \wq_i)$ are the eigenpairs of  
$(H_{\Zb} , \Gz^{-1})$, with  
the ordering $\eta_i^2 \ge \eta_{i+1}^2 $, the normalization $\wq_i^\top \Gz^{-1} \wq_i = 1$ and where $H_{\Zb} \coloneqq  \gopd^\top \, G^\top \,\Gamma_{\errorTr}^{-1}\, G \,\gopd $
 is the Hessian of the negative log--likelihood 
 $\Yb|\Zb \sim \Gauss( G \,\gopd , \Gamma_{\errorTr} ) $. 
 Moreover, \cite[Theorem 2.3]{spantini2014optimal} implies that the distance,
 at optimality,     is given by $\dr^2( \Gzyh ,\Gzy )=   \sum_{i>r}  {\rm ln}^2(\,1+\eta_i^2\,)$ 
and that the minimizer  is unique if the
first $r$  eigenvalues of $(H_{\Zb} , \Gz^{-1})$ are distinct.	
Now let $(\lambda_i , q_i)$ be defined as in Theorem \ref{approxCovUpdate-qoi},
with $\lambda_i>0$, and
 let $\Sigma\succ 0$ be the covariance matrix of the joint 
distribution of $\Yb$ and $\Zb$, i.e., 
\begin{equation} \label{eq:covJointDataQoI}
\Sigma = \begin{pmatrix}
 \Gy & G \,\Gpr\,\gop^\top \\
 \gop\,\Gpr\,G^\top & \Gz
\end{pmatrix}.	
\end{equation}
By Lemma \ref{lem:schur}[part \ref{lemSchurMain}] applied to \eqref{eq:covJointDataQoI}, we know
that $(\,\lambda_i (1 - \lambda_i)^{-1} \, , \gop\,\Gpr\,G^\top\,q_i)$ are 
eigenpairs of  
$(H_{\Zb} , \, \Gz^{-1})$.
Moreover, by Lemma \ref{lem:schur}[part \ref{lemSchurLI}] we know that 
we can always write a maximal set of 
linearly independent 
eigenvectors of $(H_{\Zb} , \, \Gz^{-1})$, associated with nonzero 
eigenvalues, as $(\gop\,\Gpr\,G^\top\,q_i)$.
Thus, since $f(\lambda)=\lambda (1 - \lambda)^{-1}$ is a decreasing function of $\lambda$
as $\lambda \downarrow 0$, we must have 
$\eta_i^2 = \lambda_i (1 - \lambda_i)^{-1}$ and we can assume, without loss
of generality, that 
$\wq_i = \alpha \, \gop \,\Gpr \,G^\top\, q_i$
for some real $\alpha>0$.
Given the normalizations $\wq^\top \Gz^{-1} \wq = 1$ and
$q_i^\top (\,G\, \Gpr \, \gop^\top \, \Gz^{-1} \,\gop \, \Gpr\, G^\top\,) q_i = 1$, 
it follows that $\alpha=1$.
Simple algebra then leads to \eqref{minimizer_theorem_covgoal} and 
\eqref{eq:error_estimate_covgoal}.
Notice,  that $\lambda_i>0$ and $f(\lambda_i)>0$ 
imply $\lambda_i<1$. 
This property will be useful when proving Lemma \ref{cor:sqrtGyz}.
\hfill $\square$
\smallskip

We now state a standard result that will be used in proving Lemma \ref{lem:comparisonApprox}.
\smallskip

\begin{theorem}[Cauchy Interlacing Theorem (e.g., \cite{kressner2014indefinite,bhatia2013matrix})] 
\label{thm:Cauchy}
Let $A,B\in\re^{n\times n}$ be symmetric matrices with $B\succ 0$ and let
$\gamma_1 \ge \gamma_2  \ge \cdots \ge \gamma_n$ be the 
eigenvalues of $(A,B)$.
For any $P\in\re^{n\times p}$, with $p\le n$ and 
full column-rank, let 
$\mu_1 \ge \mu_2 \ge \cdots \ge \mu_p$ be the eigenvalues of 
$(P^\top A \, P \, , \, P^\top B \, P )$.  Then:
\begin{equation} \label{genInterlaceIneq}
 \gamma_k \ge \mu_k  \ge \gamma_{n-p+k}, \qquad k = 1,\ldots,p.	
\end{equation}
\end{theorem}
\smallskip

{\bf Proof of Lemma \ref{lem:comparisonApprox}.} 
The first inequality in	\eqref{eq:comparisonDistances}
follows from the optimality statement of Theorem  \ref{approxCovUpdate-qoi} since
$ \Gzysub\in {\Mc}_r^{\Zb}$.
The second inequality in \eqref{eq:comparisonDistances} follows from
the Cauchy interlacing theorem (see Theorem \ref{thm:Cauchy}).
Let $\gamma_1 \ge \gamma_2  \ge \cdots \ge \gamma_n \ge 1$ be the
eigenvalues of $(\, \Gposh \, , \, \Gpos \,   )$ and
$\mu_1 \ge \mu_2 \ge \cdots \ge \mu_p$ be the
eigenvalues of $(\, \Gzysub \, , \, \Gzy \, )=(\, \gop\,\Gposh\,\gop^\top \, ,
\,\gop\,\Gpos\,\gop^\top \,)$, where $\gop$ is a full row-rank matrix.
Then, by Theorem \ref{thm:Cauchy},
\begin{equation}
\gamma_k \ge \mu_k  \ge   1, \qquad k = 1,\ldots,p.	
\end{equation}
In particular, since $\ln^2(x)$ is monotone increasing on $x > 1$, we have:
\begin{equation}
\dr( \, \Gzy \, , \, \Gzysub \,) = \frac{1}{2}\,\sum_k \ln^2(\mu_k) \le
\frac{1}{2}\,\sum_k \ln^2(\gamma_k) \le \dr(\, \Gpos \, , \, \Gposh \,),
\end{equation}
where clearly $\sum_{k>p} \ln^2(\gamma_k)\ge 0$.
\hfill $\square$
\medskip

The following two lemmas will be used in proving Lemma \ref{lem:convergHell}.
\smallskip

\begin{lemma}\label{lem:determ}
If $\Gamma_1 \succeq \Gamma_2 \succ 0$, then $|\Gamma_1|\ge|\Gamma_2|$
\end{lemma}
\smallskip

\begin{proof}
If $\Gamma_1 \succeq \Gamma_2$, then there exists a $S\succeq 0$ such that
$\Gamma_1 = \Gamma_2 + S$. Thus,
$	|\Gamma_1|\,|\Gamma_2|^{-1} = |I+\Gamma_2^{-1/2}\,S\,\Gamma_2^{-1/2}|\ge 1.$
\end{proof}
\medskip

\begin{lemma} \label{lem:boundVariance}
Let $X\sim\Gauss(\mu,\Sigma)$ and $Y\sim \Gauss(0,\Gamma)$ with $\Gamma\succeq\Sigma\succ 0$.
Let $g$ be a measurable real-valued function such that
\begin{equation}
		\Ex[|g|^{2+\alpha}(Y)]<\infty
\end{equation}	
for some $\alpha>0$. Then,
\begin{equation}
		\Ex[g^{2}(X)]\le
		\frac{|\Gamma|^{1/2}}{|\Sigma|^{1/2}}\,
		\exp \left( \frac{\mu^\top \Gamma^{-1} \mu}{\alpha} \right)\,
		\Ex[ |g|^{2+\alpha}(Y)]^{ 1/(1+\alpha/2) }
\end{equation}
\end{lemma}
\smallskip

\begin{proof}
Let $f_X$ and $f_Y$ be the densities of $X$ and $Y$, respectively, and $M_Y$ the
moment generating function of $Y$.
Since we have $\Sigma^{-1}=\Gamma^{-1}+S$ for some $S\succeq 0$, it follows that for
all $x$,
\begin{equation}
	 f_X(x)\le K\,\exp(   \mu^\top \Gamma^{-1} \, x  )\,f_Y(x),
\end{equation}	
where $K\coloneqq |\Gamma|^{1/2} \, |\Sigma|^{-1/2}\,\exp(-\mu^\top\,\Gamma^{-1}\,\mu/2)$.
Now we use H{\"o}lder's inequality, with 
$p=1+\alpha/2$ and $q=p/(p-1)$ so that $1/p+1/q=1$, to obtain:
\begin{eqnarray}
	\Ex[g^2(X)] & \le & K\,\Ex[g^2(Y)\,\exp(   \mu^\top \Gamma^{-1} \, Y  )] \nonumber \\
	& \le & K\,( \Ex[|g|^{2p}(Y)] )^{1/p}\,( \Ex[\exp(  q\, \mu^\top \Gamma^{-1} \, Y  )] )^{1/q} 
	\nonumber \\
	& = & K\,( \Ex[|g|^{2p}(Y)] )^{1/p}\,M_Y^{1/q}(q\,\Gamma^{-1}\mu) \nonumber \\
	& = & K\,( \Ex[|g|^{2p}(Y)] )^{1/p}\, \exp( q\, \mu^\top \Gamma^{-1} \mu/2 ) \nonumber \\
	& = & |\Gamma|^{1/2} \, |\Sigma|^{-1/2}\, ( \Ex[|g|^{2p}(Y)] )^{1/p} \,
	\exp( (q-1) \, \mu^\top \Gamma^{-1} \mu/2 ) \nonumber \\
	& = & |\Gamma|^{1/2} \, |\Sigma|^{-1/2}\, 
	\exp(  \mu^\top \Gamma^{-1} \mu/\alpha )
	( \Ex[|g|^{2+\alpha}(Y)] )^{1/(1+\alpha/2)},  \nonumber
\end{eqnarray} 
where we used the fact that $M_Y(t)=\exp( t^\top\,\Gamma\,t/2)$
since $Y\sim \Gauss(0,\Gamma)$.
\end{proof}
\medskip

{\bf Proof of Lemma \ref{lem:convergHell}.} 
By \cite[Lemma 7.14]{dashti2013bayesian} we have:
\begin{equation}
\left \vert  \Ex_{\mzy}[g] - \Ex_{\mzya}[g]  \right \vert \le
2\,\sqrt{\,\int | g |^2  \, (  \pizy + \ \pizya) } \, \,d_{\rm Hell}(\mzy,\mzya)	
\label{eq:boundHell1}	
\end{equation}
where $\pizy$ and $\pizya$ are, respectively, the densitites of $\mzy$ and
$\mzya$ with respect to the Lebesgue measure.
Now notice that $\Gzy \preceq \Gz$ as well as $\Gzyopt \preceq \Gz$.
Thus, by Lemma \ref{lem:boundVariance}, we have:
\begin{equation}
 \Ex_{\mzy}[|g|^2]	+  \Ex_{\mzya}[|g|^2] \le 
 2\,\frac{|\Gz|^{1/2}}{|\Gzy|^{1/2}}\,
		\exp \left( \frac{1}{\beta-2}\,\| \muzy \|_{\Gz^{-1}}^2 \right)\,
		\Ex_{\nuz}[ |g|^{\beta}]^{ 2/\beta }
\end{equation}
where we used the fact that $|\Gzyopt|\ge |\Gzy|$ since $\Gzyopt \succeq \Gzy$ 
(see Lemma \ref{lem:determ}).
Thus, \eqref{eq:boundHell} follows from simple algebra.  \hfill $\square$
\medskip

\begin{lemma} \label{lem:posGamma}
Let $M\coloneqq A(I-B B^\top)A^\top \succ 0$ for a  pair of
compatible matrices $A,B$, and let $P$ be the orthogonal projector onto the
range of $A^\top$. Then $C \coloneqq I - P B B^\top P \succ 0$ and $M=A C A^\top$.	
\end{lemma}
\smallskip

\begin{proof}
Since $M\succ 0$, $A^\top$ must be full column rank.
Thus, by definition,  $P=A^\top (A A^\top)^{-1} A$ and $P A^\top = I = A P$.	
Hence  $M=A C A^\top$. Now let $Q\coloneqq I -P$ and notice that $P Q = 0$ and
$C Q = Q$. Thus, for $z \neq 0$,
$
\langle C z , z \rangle = \langle C P z , Pz \rangle + 
\langle  Q z , Q z \rangle = \langle C P z , Pz \rangle + \| Q z\|^2
$.
In particular,
$\langle C P z , Pz \rangle = \langle P C P z , z \rangle = 
\langle M (A A^\top)^{-1} A z , (A A^\top)^{-1} A z \rangle \ge 0$ and it is
zero only if $P z = 0$, in which case $Q z \neq 0$ and $\| Q z\|>0$.
Thus $C \succ 0$.
\end{proof}
\medskip

{\bf Proof of Lemma \ref{cor:optGoalApproxCov}.} 
We first need to show the equivalence $\Gzyopt \equiv \gop\, \Gposh^* \, \gop^\top$, where
$\Gposh^*$ is defined in \eqref{eq:goalOrientedGpos}.
Notice that
$\gop\, \Gposh^* \, \gop^\top = 
\Gz - \sum_i \, \lambda_i \, v_i \, v_i^\top $ with 
$v_i \coloneqq \gop \,\Gprs\,\Pi \,\Gprs^\top \,G^\top q_i=
\gop\,\Gpr\,G^\top\,q_i$ since $\Pi$ is a projector
onto the rowspace of $\gop \,\Gprs$. 
The desired equivalence follows by comparison with 
\eqref{minimizer_theorem_covgoal}.
In particular, it follows that $\gop\, \Gposh^* \, \gop^\top \succ 0$.
Thus, in order to show that $\Gposh^*$ is in the feasible set of
\eqref{eq:goalOrientedForstner} it remains to prove that $\Gposh^*\in \Mc_r$.
Clearly, it just suffices to show that $\Gposh^* \succ 0$.
Notice that 
$\Gzyopt = \gop\,\Gprs ( I - \Pi B B^\top
\Pi  ) \Gprs^\top\,\gop^\top \succ 0 $ where
$B B^\top \coloneqq \Gprs^\top \,G^\top \sum_{i=1}^r q_i\,q_i^\top G\, \Gprs$.
Thus, we can apply Lemma \ref{lem:posGamma} with $M \coloneqq \Gzyopt$, 
$A \coloneqq \gop\,\Gprs$, 
$C \coloneqq I - \Pi B B^\top
\Pi$, and get $C \succ 0$.
In particular, this shows that $\Gposh^* = \Gprs \,C\,\Gprs^\top \succ 0$ and thus
$\Gposh^*$ is in the feasible set of \eqref{eq:goalOrientedGpos}.
Optimality of $\Gposh^*$ then follows almost immediately.
By Theorem \ref{approxCovUpdate-qoi}:
	 	\begin{equation} \label{eq:boundGoalGpos}
	 	\dr( \Gzy , \gop\, \Gposh^* \, \gop^\top ) =	
	 	\dr( \Gzy , \Gzyopt ) \le  
	 	\dr( \Gzy , \widetilde{\Gamma} )
	 	\qquad \forall\, \widetilde{\Gamma} \in  {\Mc}_r^{\Zb}. 
		\end{equation}	
In particular, we can consider $\widetilde{\Gamma}$ of the form 
$\widetilde{\Gamma}=\gop \,\Gamma\, \gop^\top$ for $\Gamma\in {\Mc}_r$.
Notice that $\widetilde{\Gamma}\succ 0$ 
since $\gop$ is assumed to be full row-rank. 
This shows optimality of  $\Gposh^*$ according to
\eqref{eq:goalOrientedForstner}.
 \hfill $\square$
\medskip

\medskip

{\bf Proof of Lemma \ref{cor:sqrtGyz}.} 
We first provide  an explicit square root
factorization of $\Gposh^*$, defined in \eqref{eq:goalOrientedGpos} 
(Lemma  \ref{cor:optGoalApproxCov}),  as 
$\Gposh^*=\Gposhs^* \,(\Gposhs^*)^\top$ for some
matrix $\Gposhs^*$.
We claim that
\begin{equation} \label{eq:sqrtGoalGpos}
	\Gposhs^* = \Gprs\,\left(\,\sum_{i=1}^r\,
	(\sqrt{1-\lambda_i}-1)\,\bar{q}_i\,\bar{q}_i^\top + I\, \right)
\end{equation}
where $I$ is the identity matrix.
and the $(\bar{q}_i)$ are
defined in \eqref{eq:formSqrtGyz}.
First of all, notice that \eqref{eq:sqrtGoalGpos} is well defined since
$1>\lambda_i>0$ for all $i=1,\ldots,r$ (see the proof of Theorem \ref{approxCovUpdate-qoi}).
One can verify that \eqref{eq:sqrtGoalGpos} is indeed a valid
square root of $\Gposh^*$. 
They key observation is that 
$\bar{q}_i = \Gprs^{-1} \widetilde{q}_i$ and that 
 the vectors $(\widetilde{q}_i)$ are $\Gpr^{-1}$-orthogonal, i.e.,
$\widetilde{q}_i^\top \, \Gpr^{-1} \,\widetilde{q}_j = \delta_{ij}$.
To see this, consider the following identities:
\begin{equation}
	\widetilde{q}_i^\top \, \Gpr^{-1} \,\widetilde{q}_j =
q_i^\top\,G\,\Gprs\, \Pi\,\Gprs^\top\,\Gpr^{-1}\, 
\Gprs\,\Pi \,\Gprs^\top \,G^\top q_j=
q_i^\top\,G\,\Gprs\, \Pi \,\Gprs^\top \,G^\top q_j. 
\end{equation}
Since 
$\Pi = \Gprs^\top\,\gop^\top\,\Gz^{-1}\,\gop\,\Gprs$, it must be
that $\widetilde{q}_i^\top \, \Gpr^{-1} \,\widetilde{q}_j =
q_i^\top \,G\,\Gpr\,\gop^\top\, \Gz^{-1}\,\gop\,\Gpr\,G^\top\,q_j=\delta_{ij}$,
for the $(q_i)$ are the generalized eigenvectors of the pencil 
\eqref{pencilGoal1}, properly normalized.
Now notice that $\Gzyopts =  \gop \, \Gposhs^*$ and thus
\eqref{eq:goalOrientedGpos} implies that 
 $\Gzyopts \,\Gzyopts^\top = \Gzyopt$.
 \hfill $\square$
\medskip

 {\bf Proof of Theorem \ref{thm:mean_approx_lowrank}.} 
 By applying \cite[Theorem 4.1]{spantini2014optimal} to the linear Gaussian model 
 defined in Lemma \ref{lem:linearModelQoi}, we know that a minimizer of \eqref{eq:optimMean} is given by:
 $A^* = \sum_{i=1}^r \,\eta_i (1+\eta_i^2)^{-1}\, \widehat{q}_i \widehat{v}_i^\top$,
where $(\eta_i^2, \widehat{q}_i)$ are 
eigenpairs of 
$(H_{\Zb}, \Gz^{-1})$ with normalization $\widehat{q}_i^\top \Gz^{-1}\widehat{q}_i=1$, whereas $(\widehat{v}_i)$ are eigenvectors of 
$( G \, \gopd \, \Gz \, \gopd^\top \, G^\top , \Gamma_{\errorTr})$ 
with normalization $\widehat{v}_i^\top  \Gamma_{\errorTr} \, \widehat{v}_i = 1$. 
Moreover, \cite[Theorem 4.1]{spantini2014optimal} tells us that the Bayes risk associated with the minimizer $A^*$ can be written as:
$\mathbb{E}[ \, \| \,  A^* \, \Yb - \Zb \, \|^2_{\scalebox{.6}{$\Gzy^{-1}$}} \,  ]  = \sum_{i>r} \, \eta_i^2 + n$,
where $n$ is the dimension of the parameter space.
The fact that the vectors $(\widehat{q}_i)$ can be written as 
$\widehat{q}_i =  \gop\,\Gpr \, G^\top\,q_i$
for $\eta_i^2>0$ was proved in Theorem \ref{approxCovUpdate-qoi}.
Furthermore, in the proof of Theorem \ref{approxCovUpdate-qoi}  
we showed that $\eta_i^2 = \lambda_i (1 - \lambda_i)^{-1}$. 
Using the latter expression we can rewrite the minimizer as
$A^* = \sum_{i=1}^r \sqrt{\lambda_i\,(1-\lambda_i)}\, \widehat{q}_i \widehat{v}_i^\top$.
If  $(\widehat{v}_i)$ are  eigenvectors of  
$( G \, \gopd \, \Gz \, \gopd^\top \, G^\top , \Gamma_{\errorTr})$, 
then they must also be   eigenvectors of
$( G\, \Gpr \, \gop^\top \, \Gz^{-1} \,\gop \, \Gpr\, G^\top \, ,\, \Gy )$. 
In particular, we can set $\widehat{v}_i = \alpha \, q_i$ for some real $\alpha>0$. 
Given the normalizations  
$q_i^\top  \,G\, \Gpr \, \gop^\top \, \Gz^{-1} \,\gop \, \Gpr\, G^\top \,q_i = 1$ and  
$\widehat{v}_i^\top  \Gamma_{\errorTr} \, \widehat{v}_i = 1$, it must be $\alpha = \lambda_i^{1/2}\,(1-\lambda_i)^{-1/2}$.
Simple algebra then leads to \eqref{minimizerMeanLowRank}. \hfill $\square$
\medskip

\bibliographystyle{siam}

\bibliography{linearGoalBib}

\end{document}